\documentclass[twocolumn,amsmath,longbibliography,amssymb,superscriptaddress]{revtex4-2}
\usepackage{graphicx}
\usepackage{dcolumn}
\usepackage{hyperref}
\usepackage{bm}
\usepackage{amsmath,amsfonts,amssymb,amsthm,verbatim}
\usepackage[ansinew]{inputenc}
\usepackage[usenames]{pstricks}
\usepackage{color}
\usepackage{epsfig}
\usepackage{MnSymbol}
\usepackage{bbold}
\usepackage{dsfont}
\usepackage{physics}

\newcommand{\be}{\begin{equation}}
\newcommand{\ee}{\end{equation}}
\newcommand{\bea}{\begin{eqnarray}}
\newcommand{\eea}{\end{eqnarray}}

\newcommand{\bi}{\begin{itemize}}
\newcommand{\ei}{\end{itemize}}
\newcommand{\bc}{\begin{center}}
\newcommand{\ec}{\end{center}}

\begin{document}

\title{Quantization of topological indices in critical chains at low temperatures}

\author{Oleksandr Balabanov}
\affiliation{Department of Physics, Stockholm University, AlbaNova University Center, SE-106 91 Stockholm, Sweden}

\author{Carlos Ortega-Taberner}
\affiliation{Department of Physics, Stockholm University, AlbaNova University Center, SE-106 91 Stockholm, Sweden}
\affiliation{Nordita, KTH Royal Institute of Technology and Stockholm University, SE-106 91 Stockholm, Sweden}

\author{Maria Hermanns}
\affiliation{Department of Physics, Stockholm University, AlbaNova University Center, SE-106 91 Stockholm, Sweden}
\affiliation{Nordita, KTH Royal Institute of Technology and Stockholm University, SE-106 91 Stockholm, Sweden}

\begin{abstract}

Various types of topological phenomena at criticality are currently under active research. In this paper we suggest to generalize the known topological quantities to finite temperatures, allowing us to consider gapped and critical (gapless) systems on the same footing. It is then discussed that the quantization of the topological indices, also at critically, is retrieved by taking the low-temperature limit. This idea is explicitly illustrated on a simple case study of chiral critical chains where the quantization is shown analytically and verified numerically. The formalism is also applied for studying robustness of the topological indices to various types of disordering perturbations.

\end{abstract}

\maketitle

\section{Introduction}

Topological phases of non-interacting quantum matter  \cite{kane_topological_2005} have attracted growing attention in the last decades. These phases are identified by the so-called topological invariants capturing some key properties of the systems. Quantum symmetries play an important role in the classification of topological matter~\cite{RuyReview} and it is common to group Hamiltonians by their response to some symmetry transformations. 
The most basic classification scheme is based on the Altland-Zirnbauer symmetry classes in Random Matrix Theory \cite{AZ1997}.
Studying the response under time-reversal, particle-hole and chiral symmetry, one finds that there exist 10 symmetry classes, the resulting classification scheme is often referred to as  ten-fold way ~\cite{schnyder_classification_2008,doi:10.1063/1.3149495}. In this paper, we focus on one-dimensional (1D) systems protected by a \textit{chiral symmetry}, commonly characterized by the winding number or quantized Zak phase~\cite{Resta_polarization_1992,KingSmith_polarization_1993,Vanderbilt_polarization_1993,Resta,Watanabe_polarization_2018}.

One of the hallmarks of 1D topological chiral phases is the existence of robust zero-energy edge modes and their connection to the topological indices via the ``bulk-edge correspondence~\cite{RuyReview}. 
At critical points the zero-energy edge modes are expected to hybridize and mix with the bulk, in this way facilitating the topological phase transitions. Nevertheless, it has been shown that critical systems in 1D can also host robust edge modes \cite{Kestner,Cheng, Fidkowski,Sau,Kraus,Keselman1,Iemini,Lang,Montorsi,Ruhman,Jiang,Zhang,Parker,Keselman2}. 
The topological origin of these edge states was first pioneered by Verresen {\it et al.} in Ref.~[\onlinecite{verresen2018}]. 
This work provided an important intuition explaining existence of the robust edge modes in critical Majorana chains, later been extended to all nontrivial Altland-Zirnbauer symmetry classes and multi-band systems in 1D~\cite{BEH}. 
Recently, the formalism was put into a larger framework of ``symmetry-enriched quantum criticality" \cite{verresenPRX} where the localized and topologically robust edge modes are implied by the presence of nonlocal symmetry operators. 
An alternative protocol for classifying critical systems was put forward in Ref.~[\onlinecite{verresen2020}]. 
Here the conventional topological indices were generalized to criticality by excluding all gapless points from the corresponding standard expressions.
Other aspects of topology in critical systems have been studied in a variety of papers, see for instance Refs.~\cite{Jones, Duque, Kumar1, Kumar2, Tantivasadakarn, Zou2021}.

The primary goal of this paper is to contribute to the effort of developing a unified classification scheme. 
Generalizing topological indices to gapless systems is neither straightforward nor unique, due on the one hand to the degeneracy of the ground state in the presence of gapless modes and on the other hand by the intrinsic fragility of critical points. 
Here, we present a conceptually novel view on the generalization of conventional topological invariants to critical systems, which leads to new theoretical and practical insights. The idea is to consider critical systems at finite temperatures: By looking at mixed thermal states we avoid the problems that originate from dealing with a degenerate ground state. The quantization of the proposed topological indices may then be retrieved by taking the low temperature limit. This can be rigorously shown for gapped topological phases, but also for some simple examples of critical systems. 

By considering different finite-temperature quantities that all converge (in presence of a gap) to the conventional topological invariant in the low temperature limit, one may design different finite-temperature generalizations of the same topological index. 
That may lead to mutually-independent classifications of critical systems. 

We here discuss two independent finite-temperature generalizations motivated by some physical intuition and practical tasks. In the first case we generalize the conventional winding number by representing it in terms of the thermal correlation matrix, which we then compute at criticality and in the low temperature limit. For translationally invariant critical systems, we recover the result from Ref.~[\onlinecite{verresen2020}] but from a conceptionally different standpoint, opening new directions for exploration. One of the possible advantages of our method, for example, is that it can be extended to systems lacking translational invariance, in contrast to Ref.~[\onlinecite{verresen2020}].
In the second generalization,  we consider a finite-temperature analogue of the Zak phase, which was first introduced for studying gapped systems at thermal-equilibrium~[\onlinecite{Bardyn2018}].  
Using their results derived for the gapped case, we adapt them to critical systems and show quantization of the finite-temperature Zak phase in the low temperature limit at critically. 

For both examples described in this paper we obtain expressions in position space. This allows us to study the robustness of the topological phases at criticality against disorder. However, since the critical points themselves are fragile, this requires a careful analysis. We first consider the effect of typical uncorrelated disorder. This disorder might open a gap, and move the system away from the critical point. For this reason we also study a special type of disordering perturbations for which the gap always remains closed. Our results indicate that the topological invariants considered here are, in fact, much more robust against the criticality-preserving disordering perturbations than to uncorrelated disorder.

The paper is organized as follows. In Sec. II the winding number is generalized to critical chiral systems in 1D by employing the thermal correlation matrix. We explicitly show its quantization in the low temperature limit, extend it to position space and conduct a comprehensive numeric study of its stability to various types of disordering perturbations. The alternative generalization method, based on many-body observables, is discussed in Sec.~III. We consider the Resta polarization at finite temperatures as generalization of the Zak phase: We also show its quantization at criticality in the low temperature limit, generalize it to position space and study its stability to various perturbations. The summary of our findings is given in the last section. Some technical calculations are put into appendices. 
 
\section{Topological indices (I): \\  The thermal correlation matrix}\label{sec:thermalC}

In this section we exemplify our construction of the topological indices on the chiral class of systems in 1D. In presence of a chiral symmetry, $H = - SHS^{-1}$ for some unitary operator $S$, in the diagonal basis of $S$ the Hamiltonians $H$ are constrained to have the off-diagonal form 
\begin{align}
\begin{split} 
H = \begin{pmatrix}
0 & h \\
 h^\dagger & 0
\end{pmatrix},
\end{split}
\label{eq:H1}
\end{align}
where $h = P_A H P_B$ with the sublattice projectors 
\begin{align}\label{eq:projectors}
  P_A &= (1 + S)/2  \nonumber\\
  P_B &= (1 - S)/2.
\end{align}
This property holds both in position and momentum space. In particular, also the  Bloch Hamiltonian $H(k)$, describing translationally invariant systems, has an off-diagonal form~\eqref{eq:H1}. 

We also note that according to Schur's first lemma~\cite{RuyReview} any irreducible representation of  chiral Hamiltonians must obey $S^2 = \exp(i\phi) \mathbb{1}$ for some complex phase $\phi$ and identity operator $\mathbb{1}$. The complex phase can be always absorbed into the definition of $S$ leaving us with $S = S^{-1}$. 

\subsection{Gapped systems}

Within the standard topological classification schemes~\cite{RuyReview}, the gapped Bloch Hamiltonians $H(k)$ are first flattened by a continuous deformation of the energies to $\pm 1$. The resulting flattened Hamiltonians $Q(k)$ can be formally defined through $Q(k) = [P_\text{val}(k) - P_\text{cond} (k)]$, where $P_\text{val}(k)$ and $P_\text{cond}(k)$ are projectors onto valence and conduction bands respectively. Note that the Q-matrices inherit the off-diagonal structure of the corresponding Bloch Hamiltonians $H(k)$:
\begin{align} 
\begin{split} 
Q(k) = \begin{pmatrix}
0 & q(k) \\
 q(k)^\dagger & 0
\end{pmatrix}.
\end{split}
\label{eq:Q1}
\end{align}
The chiral Q-matrix can be rewritten as 
\begin{align} 
\begin{split} 
Q(k) = -2[P_A P_\text{cond} (k) P_B + P_B P_\text{cond} (k) P_A],
\end{split}
\label{eq:Q1_prop}
\end{align}
that directly follows from  $P_\text{val}(k) = S P_\text{cond} (k) S$.

The flattened Hamiltonian $Q(k)$ defines a map from the Brillouin Zone to the space of matrices and may be classified by the first homotopy group. In the case of chiral systems in 1D the corresponding topological index is the winding number $\nu$ that can be calculated via~\cite{RuyReview}
\begin{align} 
\begin{split} 
\nu &= \frac{i}{2 \pi}\int^{\pi}_{-\pi} \, dk \, \text{Tr} \, [P_B Q(k) P_A \partial_k Q(k)]  \\
&= \frac{2i}{\pi}\int^{\pi}_{-\pi} \, dk \, \text{Tr} \, [P_B P_\text{cond}(k) P_A \partial_k P_\text{cond}(k)]  \\
\end{split}
\label{eq:nu1}
\end{align}
and is constrained to be an integer.

\subsection{Critical systems}

The winding number is \textit{per se} ill-defined at criticality: The conduction space, and therefore the Q-matrix, may in general change discontinuously at the gapless point(s) --- invalidating~Eq.~(\ref{eq:nu1}). To address this issue and smooth out the discontinuities we suggest to consider systems at finite temperatures. 

We propose to employ the thermal correlation matrix $C$ formally defined via
\begin{align} 
\begin{split} 
C &= \langle a^\dagger a \rangle = f(H),  \\
\end{split}
\label{eq:C1}
\end{align}
where $a^\dagger$ ($a$) is a creation (anihilation) operator, and $f(H) = (\exp(\beta H) + \mathbb{1})^{-1}$ is the Fermi-Dirac function. Note that one may make an analogous definition also within momentum space, $C(k) = f[H(k)]$. For gapped systems in the zero-temperature limit the equilibrium correlation \text{matrix} is $C(k) = P_\text{cond} (k)$, thus, hinting to the following generalization of Eq.~(\ref{eq:nu1}):
\begin{align} 
\begin{split} 
\tilde{\nu} &= \frac{2i}{\pi}\int^{\pi}_{-\pi} \, dk \, \text{Tr} \, [P_B C(k) P_A \partial_k C(k)].  \\
\end{split}
\label{eq:nu2}
\end{align}

At any non-zero temperature the correlation matrix $C(k)$ is continuous, differentiable and gauge-invariant, also at criticality, and therefore the generalized winding number $\tilde{\nu}$ is always a well defined quantity. For gapped systems one trivially recovers $\tilde{\nu} = \nu$ in the zero-temperature limit, suggesting to look for quantization of this index in the low-temperature limit also at criticality. In the following subsections, we prove the quantization in clean critical systems, and verify it numerically in the presence of certain types of disordering perturbations. 

While we used the winding number as an explicit example above, the same rationale of generalizing topological indices to critical systems will likely be applicable to other indices as well.
In particular, we believe that a similar construction will work for any index that can be expressed in terms of conduction band projectors.

\subsection{Quantization of the generalized winding number at criticality} \label{WCQ}

The generalized winding number $\tilde{\nu}$ taken for gapped systems at low temperatures is identical to the conventional winding number $\nu$ and therefore quantized. We now show its quantization also for critical systems. 

The key idea is to divide the momentum space into two parts. 
Let us start by considering a fixed low temperature. 
In any region of momentum states where the temperature scale is sufficiently smaller than the momentum-dependent gap, the correlation matrix behaves analogously to a gapped system at zero temperature, in particular $C(k) \simeq P_{\text{cond.}}(k)$. This region is called $\mathcal{P}_1$ in the following. The remaining piece $\mathcal{P}_2$ consists (for sufficiently low temperatures) of intervals surrounding each of the gapless points. 
When lowering the temperature, the intervals $\mathcal{P}_2$ shrink until becomming infinitesimal in the zero-temperature limit. 

It follows that the low temperature winding number may be represented by two contributions 
\begin{align} 
\begin{split} 
\tilde{\nu} &=  \tilde{\nu}_1 + \tilde{\nu}_2,\\
\end{split}
\label{eq:winding_k_parts}
\end{align}
where
\begin{align} 
\begin{split} 
\tilde{\nu}_1 &= \frac{2i}{\pi} \int_{k \in \mathcal{P}_1} \, dk \,  \text{Tr} \, [P_B P_{\text{cond.}}(k)  P_A  \frac{d}{dk} P_{\text{cond.}}(k)],   \\
\end{split}
\label{eq:winding_k_parts1}
\end{align}
and 
\begin{align} 
\begin{split} 
\tilde{\nu}_2 = \frac{2i}{\pi} \int_{k \in \mathcal{P}_2} \, dk \,  \text{Tr} \, [P_B C(k)  P_A  \frac{d}{dk} C(k)].   \\
\end{split}
\label{eq:winding_k_parts2}
\end{align}

In the limit of zero temperature, the second contribution $\tilde{\nu}_2$ becomes negligible, because the correlation matrix $C(k) = f[H(k)]$ is a well behaved function and the region $ \mathcal{P}_2$ becomes arbitrary small, see Appendix \ref{app:nu2} for more details.
The main contribution ${\tilde\nu}_1$ is the same as $\nu$ given in Eq.~(\ref{eq:nu1}) but obtained by excluding an infinitesimal neighborhood of the gapless point(s). It matches the index introduced in Ref.~[\onlinecite{verresen2020}] where it is shown to be quantized to \textit{half integer values} and linked to \textit{the topological boundary modes}. It is important to stress, however, that our construction is fundamentally different from the one described in Ref.~[\onlinecite{verresen2020}] --- it is introduced via smoothing of the gapless discontinuities employing the finite temperature formalism. One immediate advantage of our approach, for example, is that it can be transferred to position space, allowing us to study stability of the generalized winding number $\tilde{\nu}$ to various types of disorder.

\subsection{The generalized winding number within real-space representation}

The conventional winding number $\nu$ can be represented within position space and be used to characterize strongly disordered gapped systems~\cite{real_nu1, real_nu2}. The corresponding real-space expression reads as follows
\begin{align} 
\begin{split} 
\nu &= - \, \mathcal{T} \, \{ \, P_B Q P_A [X, Q] \, \}  \\
\end{split}
\label{eq:winding_real1}
\end{align} 
where $Q = (P_\text{cond} - P_\text{val})$ is the flattened Hamiltonian, and $X = \sum_j j |j\rangle \langle j|$ is the position operator. Here the bulk trace per volume $\mathcal{T} \{A\}$ is defined by \begin{align} 
\begin{split} 
\mathcal{T} \{A\} = \text{Tr}_{(N,M)}[A]=\sum_{j=N}^{M} \text{Tr} \, A_{jj}/(M-N), \\
\end{split}
\label{eq:trace_per_volume}
\end{align}  
where $N$ and $M$ are some bulk cell indices.

The generalized winding number $\tilde{\nu}$, Eq.~(\ref{eq:nu2}), can be adapted to position space in an analogous way: In Appendix \ref{app:generalizedWinding} we provide a detailed derivation of the corresponding formula, but essentially we replace the derivative $\partial_k$ in Eq.~(\ref{eq:nu2}) by a commutator $-i[X, \, ]$ and the integral over momentum becomes the bulk trace per volume~\cite{real_nu1, real_nu2}:
\begin{align} 
\begin{split} 
\tilde{\nu} &= - \, 4 \, \mathcal{T} \, \{ \, P_B C P_A [X, C] \, \},  \\
\end{split}
\label{eq:winding_real2}
\end{align} 
where $C$ is the real-space correlation matrix. With this expression in place, we can now proceed to study effects of disorder.

\begin{figure}
	\centering
	\includegraphics[width=66mm]{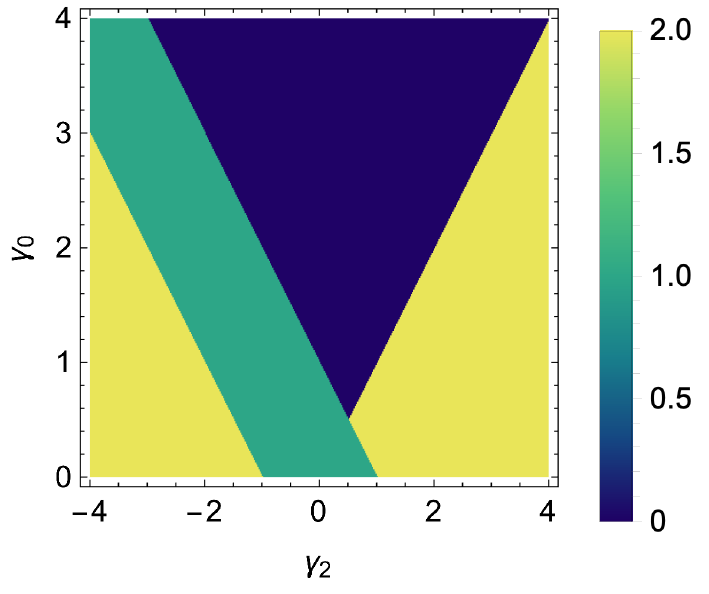}
	\caption{The conventional winding number $\nu$ of the extended SSH model considered in Sec.~IIE. Plotted for $\gamma_1 = 1$.}
\label{fig:fig1}
\end{figure}

\subsection{Stability to disorder}

Any generic disorder applied to a critical system is expected to open a gap and destroy the critical phase. Moreover, even low-amplitude disorder realizations applied at a critical point may move the system to different gapped phases making the critical points \textit{intrinsically fragile} to perturbations, even if the symmetries are preserved. The topological indices defined at criticality are, thus, not expected to be in general robust. Nevertheless, for a class of disordering perturbations that do not open up a gap and respect the symmetries, the generalized indices are anticipated to stay put.  
If so, this may be interpreted as topological protection for critical systems, albeit in a much more restricted sense than in gapped systems. In the remainder of this section, we numerically confirm this anticipation.

\subsection{Numeric test}

\subsubsection{Model}
 
To showcase our results we consider a model that is rich on topological phase transitions. The conventional SSH model is far too simple, but by adding second neighbour hopping we can obtain critical points between phases with winding numbers $\nu=(0,1)$, $\nu=(1,2)$ and $\nu=(2,0)$, as we can see in the phase diagram shown in Fig. \ref{fig:fig1}. Explicitly, the considered model is given by~\cite{COTMH}
\begin{align} 
\begin{split} 
H =  \sum_j [ \gamma_0 |A, j\rangle  & \langle B, j| + \gamma_1 |B, j\rangle \langle A, j + 1| \\
&+ \gamma_2 |B, j\rangle \langle A, j + 2|  + \text{H.c.}]\\
\end{split}
\label{bdi_model}
\end{align}
and the corresponding Bloch Hamiltonian $H(k)$ is 
\begin{align}
\begin{split} 
H(k) =& (\gamma_0 + \gamma_1 \cos(k)+ \gamma_2 \cos(2k))\sigma_x \\
&- (\gamma_1 \sin(k) + \gamma_2\sin(2 k))\sigma_y ,\\
\end{split}
\label{bdi_model_bloch}
\end{align}
where $A$ and $B$ denote the sublattices, $\sigma_x$ and $\sigma_y$ are Pauli matrices, and $\gamma_{i}$ with $i = 0, 1, 2$ are the intracell ($i = 0$) and intercell ($i = 1, 2$) hopping amplitudes. The intercell amplitudes $\gamma_1$  and $\gamma_2$ describe first and second neighbour hoppings respectively.

\begin{figure}
	\centering
	\includegraphics[width=86mm]{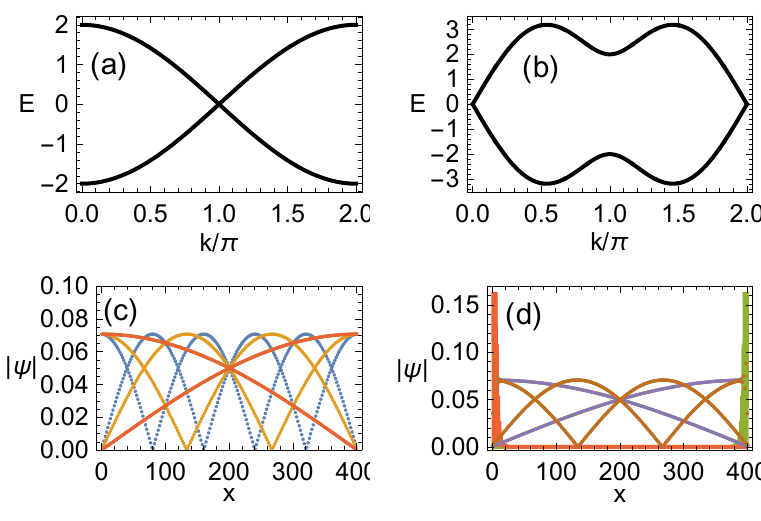}
	\caption{Energy spectrum of the Bloch Hamiltonian at the two critical points we consider in this text, (a) $\gamma_0=\gamma_1, \gamma_2=0$ and (b) $\gamma_0=\gamma_1, \gamma_2=-2\gamma_1$. (c) and (d), low-lying eigenstates of the open chain at the trivial ($\gamma_0=\gamma_1, \gamma_2=0$) and topological ( $\gamma_0=\gamma_1, \gamma_2=-2\gamma_1$) critical points, respectively.}
\label{fig:fig2}
\end{figure}

In Fig.~\ref{fig:fig2} we show the energy spectrum of the Bloch Hamiltonian $H(k)$ for two critical points at the interface between $\nu= (0,1)$ and $\nu=(1,2)$. We also show a few low-lying states of open chains at both critical points. In agreement with the theory~\cite{verresen2018, verresen2020} the critical point between $\nu=(1, 2)$ supports edge modes while the one between $\nu=(0,1)$ does not.

\subsubsection{Unperturbed chains}
First, we focus on the ideal case without any disordering perturbations. 
\begin{figure}
	\centering
	\includegraphics[width=\columnwidth]{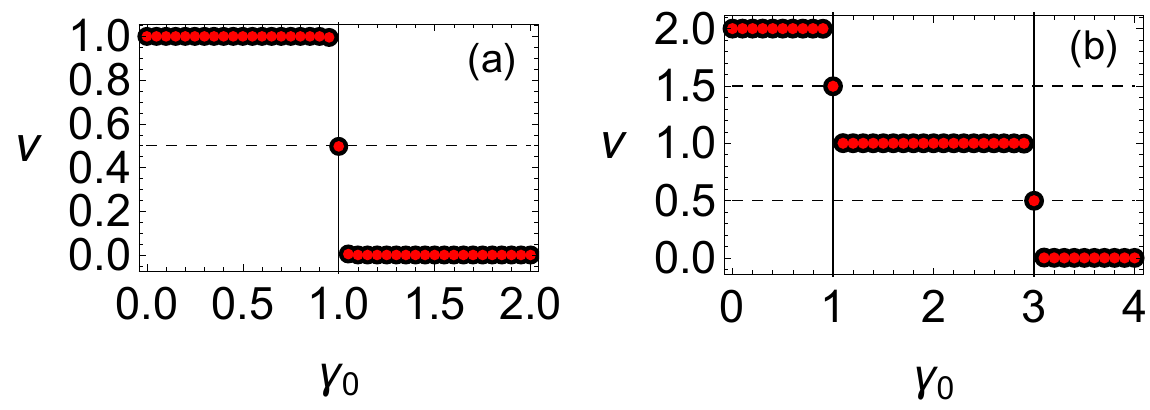}
	\caption{The generalized winding number $\tilde{\nu}$ of the extended SSH model considered in Sec.~IIE. (black dots) computed in momentum space and (red dots) computed using the position space formula. Plotted for $\gamma_1 = 1$ and (a) $\gamma_2=0$ and (b) $\gamma_2=-2$. $L=1000$ sites and temperature $\beta=100$. Dashed lines at half-integers to visualize the quantization at the critical points, signaled by vertical lines.}
\label{fig:fig3}
\end{figure}
In Fig.~\ref{fig:fig3} we plot the low-temperature generalized winding number $\tilde{\nu}$ computed within both momentum-space (black dots) and real-space (red dots) representations, Eqs.~(\ref{eq:nu2}) and~(\ref{eq:winding_real2}) respectively.
Note that they perfectly coincide, verifying the equivalence between the two expressions.

Within the gapped regions, the generalized winding number coincides with the usual 
definition~\eqref{eq:nu1}, and we obtain the values expected from the phase diagram in Fig.~\ref{fig:fig1}. 
At the phase boundaries, the generalized winding number is seen to be quantized to half integer values, in agreement with the original prediction of Ref.~[\onlinecite{verresen2020}].

\subsubsection{Disordering perturbations}
\label{sec:disorder}
We now turn our attention to a system that could not be studied with the construction in Refs.~\cite{verresen2018,verresen2020}.
Consider a disordering perturbation that maintains the chiral symmetry, denoted by $\Delta H$. The chiral symmetry allows the terms to be explicitly represented as follows  
\begin{align} 
\begin{split} 
& \Delta H  = \sum_{i,j} \left( \delta h_{i, j} | A, i\rangle \langle B, j | + \text{H.c.} \right); \\
&H'  = H + \Delta H =  \sum_{i,j}  \left( h'_{i, j} | A, i\rangle \langle B, j | + \text{H.c.} \right), \\
\end{split}
\label{eq:DH_H}
\end{align}
where $A$ and $B$ are the sublattices, and $H'$ denotes the total Hamiltonian.

\begin{figure}
	\centering
	\includegraphics[width=\columnwidth]{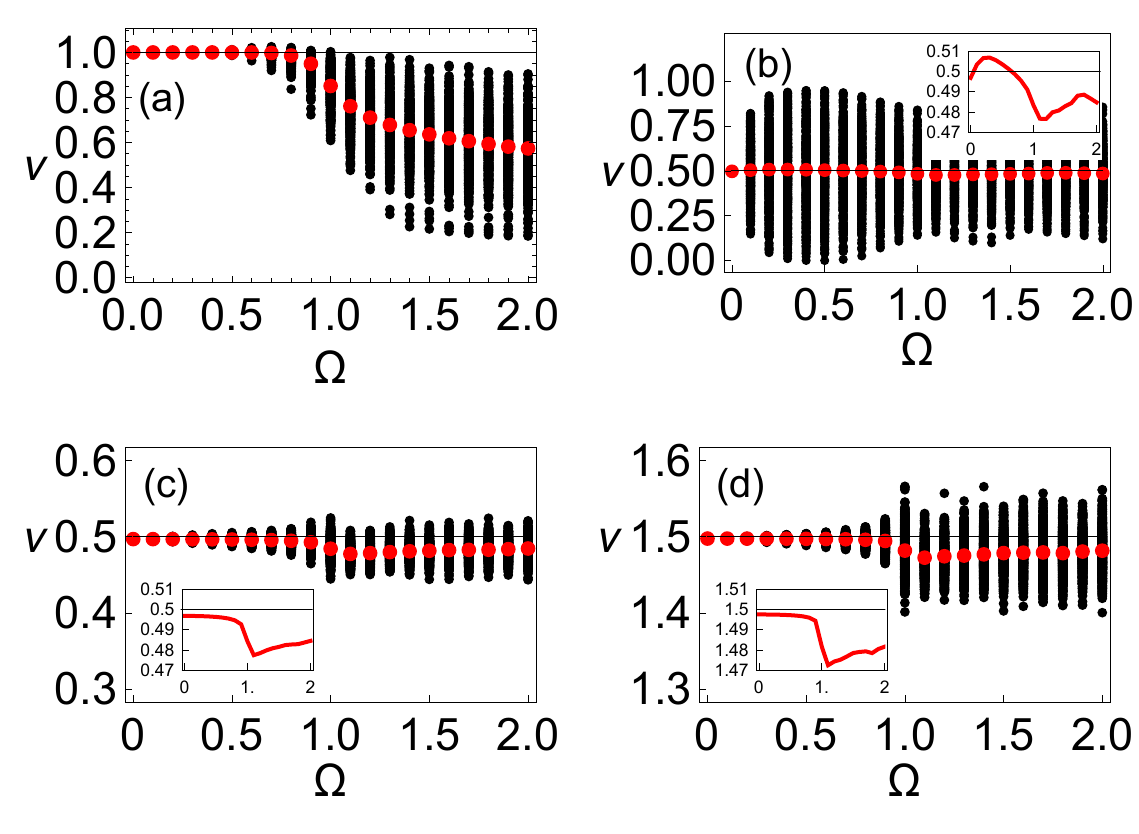}
	\caption{The generalized winding number $\tilde{\nu}$ of the disordered SSH chains. Considered for $\gamma_1 = 1$ and (a) $\gamma_0 = 0.5$, $\gamma_2 = 0$ (gapped), (b) - (c) $\gamma_0 = 1$, $\gamma_2 = 0$ (gapless), and (d) $\gamma_0 = 1$, $\gamma_2 = -2$ (gapless). Evaluated for $500$ independent disorder realizations. The plots contain the mean (black dots) and max deviation from the mean (solid intervals). The perturbation disorders $\gamma_0$ and $\gamma_1$ in (a) - (b), and preserves criticality in (c) - (d) in accordance with Sec. IIF. Note the different scale in (c) and (d) compared to (b). The insets show the deviation of the mean from the quantized value. The perturbation value at each site (cell) is uniformly selected from $[-\Omega, \, \Omega]$.  $L=1000$ sites and temperature $\beta=100$.}
\label{fig:fig4}
\end{figure}

In Fig.~\ref{fig:fig4} we study the stability of the low-temperature generalized winding number $\tilde{\nu}$ to different types of chiral disordering perturbations $\Delta H$. First we consider a perturbation corresponding to completely uncorrelated disorder: We take the onsite elements $\delta h_{i,i}$ and nearest neighbor hoppings $\delta h_{i,i-1}$  of $\Delta H$  to be randomly distributed within $[-\Omega, \, \Omega]$ with some amplitude $\Omega$ and all other elements are put to zero. In Figs.~\ref{fig:fig4} (a) and (b) we display the stability of $\tilde{\nu}$ to this perturbation for gapped and gapless chains, respectively. For the gapped case in Fig.~\ref{fig:fig4} (a), the topological index is seen to be robust to generic symmetry-preserving disorder that is in agreement with Refs.~[\onlinecite{real_nu1, real_nu2}]. 
In Fig.~\ref{fig:fig4} (b) we apply the same perturbation but to a critical system: The quantization of the winding number gets immediately destroyed by the disorder. 
This behavior is not surprising, because generic disorder opens a gap that by default kills criticality. 
Interestingly, the generalized winding number in Fig.~\ref{fig:fig4} (b) is still found to fluctuate (on average) around the quantized value for small disorder ($\Omega \lesssim 0.9)$. For larger disorder, the average deviates from $1/2$, similar to what we find later for the disordering perturbations. We, thus, conjecture that a residue of topological protection to generic symmetry-preserving disorder may still be present in critical systems. 

Critical systems crucially hinge on the absence of a gap. We, therefore, look at a class of disordering perturbations that keep the gap closed and study the effect of such perturbations on the topological number $\tilde{\nu}$ numerically. 
To exemplify our results we consider two sets of parameters: 1. $\gamma_0 = \gamma_1 = 1$, $\gamma_2 = 0$ and 2. $\gamma_0 = \gamma_1 = 1$, $\gamma_2 = -2$. Both of them correspond to a critical point, see Fig.~\ref{fig:fig3}. 

The disordering perturbations $\Delta H$ that we consider for each of the two cases are chosen to maintain the following constraint placed on the total Hamiltonian $H^\prime$:
\begin{align} 
\begin{split} 
&\text{1. $h'_{i, i - 1} = h'_{i, i}$ and $h'_{i, j} = 0$ otherwise;} \\
&\text{2. $h'_{i, i - 1} = h'_{i, i} = -\frac{1}{2} \, h'_{i, i -2}$ and $h'_{i, j} = 0$ otherwise.}\\
\end{split}
\label{eq:Cases}
\end{align}
The unperturbed Hamiltonian $H$ already fulfills this requirement for the corresponding set of parameters and we require the disordering perturbation to satisfy these conditions as well, with the corresponding amplitudes being randomly chosen from $[-\Omega, \, \Omega]$. Neither of these two relations on $H^\prime$ opens up a gap (see Appendix \ref{app:disorderingperturbations}). Note that such disordering perturbations only bind hopping amplitudes that correspond to the same cell making it locally fine-tuned, but globally uncorrelated.     

In Figs.~\ref{fig:fig4} (c) and (d) we plot the results for choices $1$ and $2$, respectively. The winding number $\tilde{\nu}$ is seen to be robust to this type of disorder [note the difference in scales compared to the effect to uncorrelated disorder in (b)], confirming our expectation for the topological numbers to be robust to criticality-preserving disordering perturbations. Note that at strong disorders the topological index $\tilde{\nu}$ starts to exhibit fragility, and deviates from the quantized value, similar to the behavior for uncorrelated disorder shown in (b). This behavior is also expected.

\section{Topological indices (II): \\ \ \ \ \  Many-body observables}

In Sec.~II we described a procedure for generalizing  topological indices to finite temperatures via representing them in terms of the thermal correlation matrix $C(k)$.

In this section we present an alternative construction for topological indices at criticality. It also lies within the finite-temperature framework but the build-up is conceptually different. The idea is to represent the conventional topological indices as expectation values of many-body observables, generalize them to finite temperatures, and then take the low-temperature limit for observing the quantization. The benefit of this approach is that it is directly linked to a measurable quantity and therefore has evident practical implications. 

In the following we discuss the generalization of the $Z_2$ invariant, the so-called  Zak phase $\phi_\text{Zak}$\cite{Zak, RuyReview}: We show quantization of the generalized Zak phase at criticality and study its robustness against different types of disordering perturbations. For simplicity we focus only on a chiral class of two-band systems in 1D. The indices beyond $Z_2$ in 1D, even the winding number $\nu$, seem to be much more challenging to address within this framework and need more thoughtful analysis. We leave this topic for future research.

\subsection{Resta polarization at finite temperatures}

A translationally invariant 1D gapped system can be formally solved by diagonalizing the corresponding Bloch Hamiltonian $H(k)$. Its eigenstates, $|n; {k}\rangle$ where $n$ is the band index, are momentum-periodic and can be used to compute the so-called Zak phase $\phi_\text{Zak}$ $-$ a Berry phase acquired by the occupied states after completing a loop within the Brillouin Zone\cite{Zak}. For discretized systems, the Zak phase is obtained as the phase of the Wilson loop, given for an isolated band by
\begin{equation}
W_{n} = \prod_i \bra{n;k_{i+1}}\ket{n;k_{i}}.
\end{equation}
The Zak phase, $\phi_\text{Zak}$, of a gapped periodic system can be linked to the so-called Resta polarization\cite{Resta} defined as
\begin{align} 
\begin{split} 
P = \frac{\phi_\text{Zak}}{2\pi} = \frac{1}{2\pi}\text{Im} \, \ln \, \langle \Psi | \hat{T} |\Psi\rangle, \ \ \ \ \ \hat{T} = \exp(i \delta k \hat{X}), \\
\end{split}
\label{eq:Resta}
\end{align}
where $|\Psi\rangle$ is the ground state, $\hat{X} = \sum_j \hat{x}_j$ is the many-body position operator ($\hat{x}_j$ is the position operator of individual particles indexed by $j$), and $\delta k = 2 \pi / L$ with system size $L$. 

In Ref.~[\onlinecite{Bardyn2018}], Bardyn \textit{et al.} generalized the Zak phase to finite temperatures by simply considering 
\begin{align}
P = \frac{\tilde{\phi}_\text{Zak} }{2\pi}  = & \frac{1}{2\pi} \Im \ln \expval{\hat{T}} \nonumber \\
=& \frac{1}{2\pi} \Im \ln \Tr[\rho \hat{T}],\label{eq:Ptemp}
\end{align}
where $\rho$ is the thermal density matrix at finite temperature.
They used this expression to study gapped systems and showed that in the low-temperature limit, one recovers the polarization of the filled bands.
Following their results, the generalized Zak phase, denoted by $\tilde{\phi}_\text{Zak}$, may be represented as
\begin{align} 
\begin{split} 
\tilde{\phi}_\text{Zak} = \text{Im} \, \ln \, \det \, [\mathbb{1} + M_T]\\
\end{split}
\label{eq:ZP}
\end{align} 
with
\begin{align} 
\begin{split} 
M_T = (-1)^{L+1}\Pi_i \, \exp(-B_{i+1}) \, U_{i+1, i},  \\
\end{split}
\label{eq:M_T}
\end{align} 
where $B_i = \text{diag}_n(\beta_{i,n})$  is a diagonal matrix constructed from the corresponding Boltzmann factors $\beta_i=\beta E_i$, for every eigenenergy $E_i$. The link matrices $U_{i+1, i}$ are defined via $[U_{ {i+1,i}}]_{n m} =  \langle n; {k}_{i+1} | m; {k}_{i}\rangle$, where $(n, m = 1; 2)$ for two-band models. Note that there is a $\pi$-shift in $\tilde{\phi}_\text{Zak}$ under a parity change of $L$.

\subsection{Gapped systems in the zero-temperature limit}

Before proceeding with critical systems it is instructive to first consider gapped systems. 
While the results of Ref.~[\onlinecite{Bardyn2018}] hold for any number of bands and without symmetry constraints, we will for sake of simplicity restrict the discussion to chiral-symmetric two-band models. 

In presence of a chiral symmetry the conventional Zak phase $\phi_\text{Zak}$ is quantized to $(\nu \text{ mod } 2)\, \pi = \{0; \, \pi\}$ and defines a $Z_2$ classification of gapped systems in 1D. In the following, we will show that for gapped two-band models the Zak phase $\tilde{\phi}_\text{Zak}$ at low temperatures, Eq.~(\ref{eq:ZP}), correctly reproduces the quantized values. 

To begin with, we notice the following identity
\begin{align} 
\begin{split} 
\det \, [\mathbb{1} + M_T] = \det M_T + \text{Tr} M_T + 1\\
\end{split}
\label{eq:ZP1}
\end{align}
valid for  $\text{dim}(M_T) = 2$, i.e. two-band models. Chiral symmetry implies that $ \det \, \exp(-B_i) = 1$ and therefore
\begin{align} 
\begin{split} 
\det M_T & = (-1)^{L+1} \det [\, \Pi_i \, \exp(-B_i) \, U_{i+1, i}]\\
& = (-1)^{L+1} \Pi_i \, \det \, \exp(-B_i) \, \det U_{i+1, i}\\
& = (-1)^{L+1} \Pi_i \, \det U_{i+1, i} = (-1)^{L+1}.\\
\end{split}
\label{eq:ZP2}
\end{align}

Thus, the expression for the Zak phase $\tilde{\phi}_\text{Zak}$ reduces to
\begin{align} 
\begin{split} 
\tilde{\phi}_\text{Zak} &= \text{Im} \, \ln \, [(-1)^{L+1}  + 1 + \text{Tr} M_T]\\
 &\simeq \text{Im} \, \ln \, [(-1)^{L+1} \, e^{\sum_j \beta_j}W_\text{cond.}] \\
& = \phi_\text{Zak},\\
\end{split}
\label{eq:ZP3}
\end{align}
where $L$ is taken to be an odd integer and $ W_\text{cond.}$ is the Wilson loop associated with the conduction band. 

The approximation in the second line of Eq.~\eqref{eq:ZP3} requires some explanation: Because of the product of link matrices in $M_T$, each element will be a sum of different terms corresponding to all paths connecting $\ket{n,k=0}$ and $\ket{m,k=2\pi}$, with any number of jumps between the bands in between. However, due to the presence of the Boltzmann term, any jump to the valence band is exponentially suppressed in the low-temperature limit. The leading contribution to the trace is therefore the term with no contribution from the upper band, which is just the Wilson loop of the conduction band, with the additional Boltzmann factors.

\subsection{Gappless systems in the zero-temperature limit}
Above, we showed how the generalized Zak phase approaches the usual Zak phase in the limit of zero temperature. 
We now proceed with discussing gapless systems. 
While they can be considered on a similar footing, they do require a more thorough analysis. We again restrict our discussion to chiral-symmetric two-band systems.  For simplicity, we assume that the system has only a single gapless point in momentum space, which without loss of generality we set at $k=0$. The generalization to multiple gapless points is straightforward. 

For a two-band chiral system, the valence state $| 1; {k} \rangle$ with energy $E_1(k)\leq 0$ and conduction state $| 2; {k} \rangle$ with energy $E_2(k)= -E_1(k)$ must obey
\begin{align} 
\begin{split} 
| 1; {k} \rangle= \frac{1}{\sqrt{2}}
\begin{bmatrix}
e^{i \phi(k) }  \\
1 
\end{bmatrix}
\ \ \ \ \ 
| 2; {k} \rangle = \frac{1}{\sqrt{2}}
\begin{bmatrix}
-e^{i \phi(k) }  \\
1 
\end{bmatrix},
\end{split}
\label{eq:states2}
\end{align}
where $\phi(k)$ is a periodic function that is continuous \text{everywhere} except possibly at $k = 0$. Note that the chiral symmetry operator is chosen as $S = \sigma_z$.

Since the eigenspace composed of $| 1; {k} \rangle$ and $| 2; {k} \rangle$ has to be periodic, there are only two scenarios at  $k = 0$:
\begin{align} 
\begin{split} 
&\text{(a) $\phi(k = 2 \pi_-) = \phi(k = 0_+) + 2 \pi n$}; \\
&\text{(b) $\phi(k = 2 \pi_-) = \phi(k = 0_+) + \pi + 2\pi n$}, \\
\end{split}
\label{eq:list}
\end{align}
where in (a) the eigenstates are continuous everywhere  and  in (b) they flip at $k = 0$.

In Appendix D we discuss in details that in (a) we recover the conventional Zak phase, $\tilde{\phi}_\text{Zak} =  \phi_\text{Zak}$, while in (b) the output is
\begin{align} 
\begin{split} 
\tilde{\phi}_\text{Zak} \approx  \left(\text{Im} \, \ln \, (-1)^{L+1}W^\text{cond.}_{\Delta<j<L -\Delta}\right) - \frac{\pi}{2}\text{sign}(\partial_{k}\phi(k)|_{0^+}),
\end{split}
\label{eq:Zak2}
\end{align}
where 
\begin{align} 
\begin{split} 
W^\text{cond.}_{\Delta<j<L -\Delta}  = \langle 2; k_\Delta | 2; k_{\Delta + 1} \rangle \langle 2; k_{\Delta + 1} | ... | 2; k_{L - \Delta} \rangle
\end{split}
\label{eq:WDelta}
\end{align}
is the Wilson loop obtained by avoiding a neighborhood of the gapless point given by $\Delta \ll L$. It is equivalent to $(\tilde{\nu} \mod 2) \, \pi$, which is quantized to half-integers for gapless phases. The second term complements this winding number making the Zak phase quantized to $0$ or $\pi$. 

\subsection{Numeric Test}

For implementing a numeric test we consider the model from Sec.~II~F.
\begin{figure}
	\centering
	\includegraphics[width=\columnwidth]{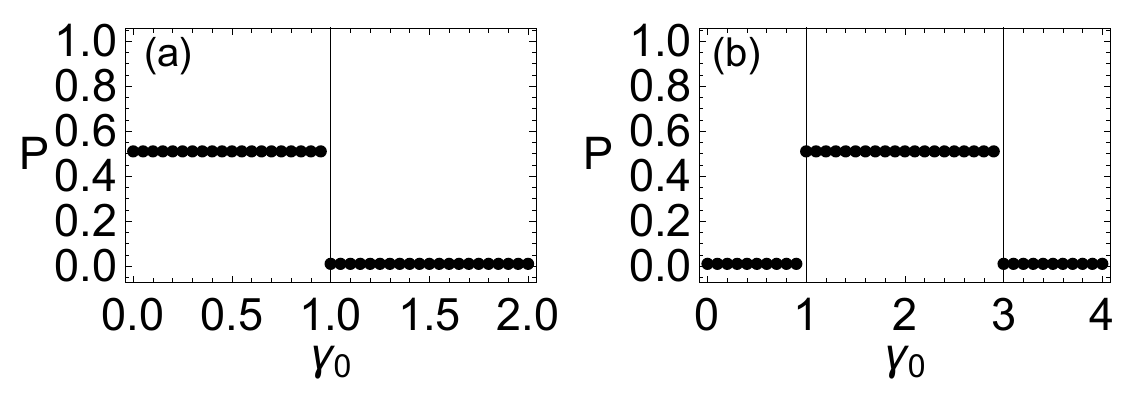}
	\caption{Polarization plotted for $\gamma_1 = 1$ and (a) $\gamma_2=0$ and (b) $\gamma_2=-2$. $L=1000$ sites and temperature $\beta=100$. Lines mark the critical points to visualize better how the Polarization takes different values for the different critical phases. The polarization is computed exactly at these critical points. }
\label{fig:fig5}
\end{figure}

\subsubsection{Unperturbed chains}

In the low-temperature regime the Boltzmann factors in Eq.~\eqref{eq:ZP} are very large for $E_n(k) < 0$ and very small for $E_n(k) > 0$ making Eq.~\eqref{eq:ZP} problematic to be applied in numerical simulations. There is an alternative expression for the finite-temperature Zak phase, cf. Ref.~[\onlinecite{Bardyn2018}], that is given by
\begin{equation}
\tilde{\phi}_\text{Zak}= \Im \ln \det[\mathbb{1}-C+C\exp(i2\pi X/L)],
\label{eq:ZakN}
\end{equation}
where $C$ is the thermal correlation matrix and $X = \sum_j j |j\rangle \langle j|$ is the position operator. Eq.~\eqref{eq:ZakN} is equivalent to Eq.~\eqref{eq:ZP} but its numerical implementation is much more straightforward. Moreover, it is given within position space and therefore it is applicable for studying disorder.

In Fig.~\ref{fig:fig5}, we plot the low-temperature polarization $P = \tilde{\phi}_\text{Zak} / 2 \pi$, for different unperturbed SSH chains. In the gapped regions, the polarization is consistent with the phase diagram in Fig.~\ref{fig:fig1} with $2 P = (\nu \text{ mod } 2)$. In agreement with our theory, the polarization is also quantized at every critical point in the figure.

\subsubsection{Disordering perturbations}

\begin{figure}
	\centering
	\includegraphics[width=\columnwidth]{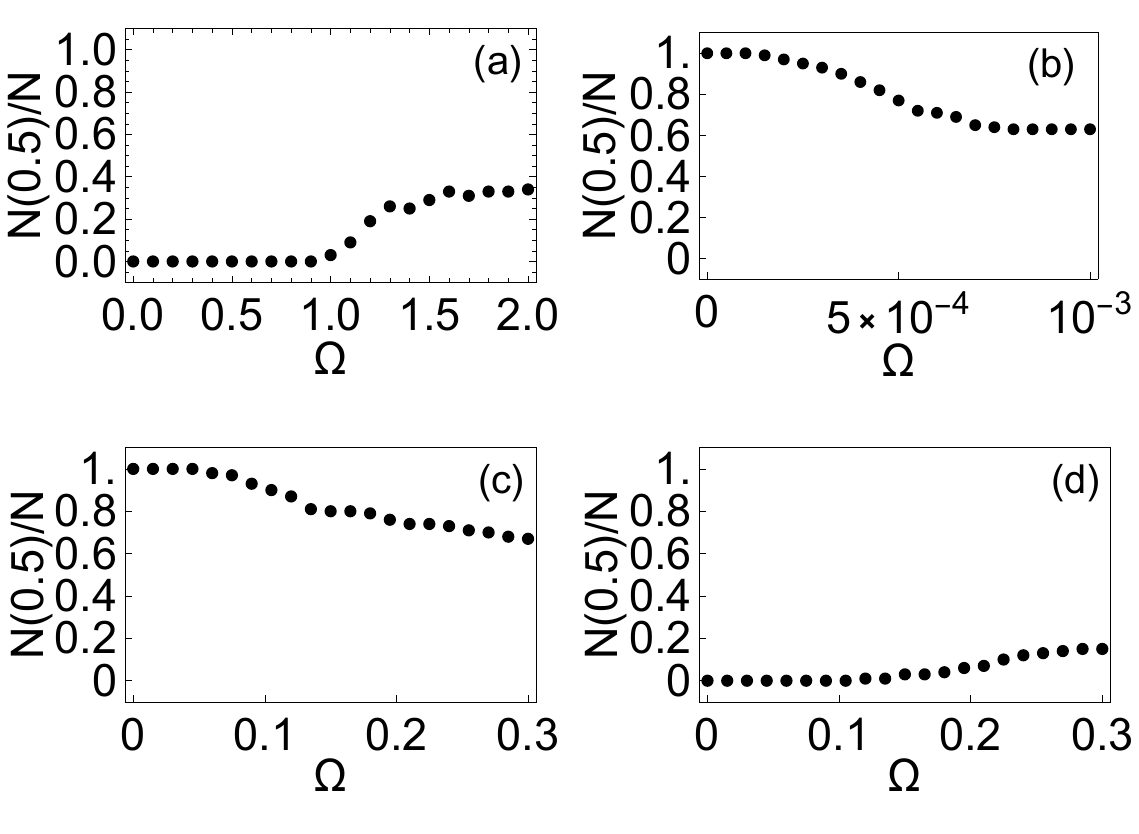}
	\caption{Polarization calculated for disordered SSH chains. Plotted for $\gamma_1 = 1$ and (a) $\gamma_0 = 0.5$, $\gamma_2 = 0$ (gapped), (b) - (c) $\gamma_0 = 1$, $\gamma_2 = 0$ (gapless), and (d) $\gamma_0 = 1$, $\gamma_2 = -2$ (gapless). Computed for 100 independent disorder realizations, we show the number of realizations with $P=1/2$ over the total. The perturbation disorders $\gamma_0$ and $\gamma_1$ in (a) - (b), and preserves criticality in (c) - (d) in accordance with Sec. III D. The perturbation value at each site (cell) is uniformly selected from $[-\Omega, \, \Omega]$.  $L=1001$ sites and temperature $\beta=100$. }
\label{fig:fig6}
\end{figure}

In Fig.~\ref{fig:fig6} we study robustness of the low-temperature polarization $P$ to the same disordering perturbations applied in Sec.~\ref{sec:disorder}. Even in the presence of disorder, the polarization is quantized to $0$ and $\pi$, which makes it hard to visualize the effect of disorder on a scatter plot. 
For this reason we plot the ratio of the disorder realizations for which $P=1/2$. 
For the gapped case, shown in Fig.~\ref{fig:fig6}(a), the polarization is robust against weak uncorrelated disorder in $\gamma_0$ and $\gamma_1$. For strong disorder some realizations are in different topological phases, which is why the ratio of $P=1/2$ deviates from 0 or 1.

The same holds also for the critical case but we find that the robustness is only present at extremely weak disorder. This can be explained by the following: The polarization is equivalent to $\arg{\expval{T}}$. In the low-temperature limit, while gapped systems have $\abs{\expval{T}} \approx 1$, critical systems scale as $\abs{\expval{T}} \propto 1/L$. Disorder makes $\abs{\expval{T}}$ deviate from its original value. Some disorder realizations can cause $\expval{T}$ to change sign, which in turn leads to an additional $\pi$ phase. For uncorrelated disorder and large systems this occurs already at extremely weak disorder. In the thermodynamic limit, robustness will be lost. 

In Fig.~~\ref{fig:fig6}(c) - (d) we study the criticality-preserving disordering perturbations from Sec.~~\ref{sec:disorder}. The effect of this disordering perturbation on the critical system is much smaller than the uncorrelated disorder [Note the difference of scales of $\Omega$ compared to \ref{fig:fig6}(b)]. 
The deviation of $\abs{\expval{T}}$ from its translationally invariant value due to disorder, $\Delta \abs{\expval{T}}$, is necessarily bounded for a particular system size and disorder strength. Therefore there is a finite range of disorder strength (for fixed system size) for which there are no sign changes due to disorder.  This allows us to conclude that the polarization is robust against certain disordering perturbations for finite system sizes.

\subsection{Experimental realization}

As pointed out in Ref.~[\onlinecite{Bardyn2018}], this generalization of the Zak phase is not equivalent to the physical polarization, and cannot be measured as by the usual current measurements. Instead, the authors proposed  a setup to measure $\expval{T}$ via interferometry. At no point in the derivation it is assumed that the system is gapped, and therefore one should be able to use the same experiment to obtain $\expval{T}$ for critical systems. However, critical systems have a disadvantage over gapped systems. The measured quantity is of order $\abs{\expval{T}}$, and therefore to improve the measurement ideally one wants to maximize $\abs{\expval{T}}$. At zero temperature, gapped systems show $\abs{\expval{T}}=1$, and for high temperatures one can even recover $\abs{\expval{T}}\simeq 1$ for small systems, as discussed in Ref.~[\onlinecite{Bardyn2018}]. 

The case of critical systems is different. Because of the discontinuity in the conduction band the Wilson loop is of order $1/L$, such that $\abs{\expval{T}}$ vanishes in the thermodynamic limit. Even for small systems $\abs{\expval{T}} << 1$, which makes measurement more difficult than in the gapped case.

\section{Summary}

In this paper we have proposed a new approach for generalizing topological invariants for gapped systems to critical systems.
The main idea is to consider the system at thermal equilibrium, and study the limit of zero temperature. 
This can be done in inequivalent ways, and we have shown two different procedures --- for the cases of the winding number and the polarization, respectively --- leading to two independent classification schemes. 
 Although, contrary to the gapped case, the generalizations of the two topological invariants are not equivalent, they correctly distinguish between different topological phases. 
 Furthermore, we have shown that under certain types of disordering perturbations that preserve criticality (i.e. do not open up a gap), the topological features are robust.

\emph{Outlook}: There are several ways one can extend the results of this paper. 
While we focus in this paper on two-band models with chiral symmetry, the same ideas can be applied to more general cases, in particular  models with many-bands,  and/or systems with other symmetries.
Furthermore the same ideas discussed here can be applied to generalize other topological invariants to critical systems, as long as the invariant can be expressed in terms of conduction-band projectors or observables. 
In particular, one can extend the analysis of Sec.~\ref{sec:thermalC} to 2D and 3D systems by generalizing the Chern number and 3D winding number in a similar way. An extension of the proposed finite-temperature protocol to periodically driven (Floquet) systems is another interesting direction for exploration.

{\bf Acknowledgements} ---
The project is supported by the Knut and Alice Wallenberg Foundation and a part of the Wallenberg Academy Fellows project.
It was also supported by the Swedish Research Council under grant no. 2017-05162. 

\appendix 

\section{The contribution $\tilde{\nu}_2$}\label{app:nu2}

Here we show that the contribution $\tilde{\nu}_2$ in Eq.~(\ref{eq:winding_k_parts2}) is negligible in the limit of zero temperature. Without loss of generality let us consider the case with just one gapless point, say at $k = 0$. Thus, our goal is to prove that in the low-temperature limit
\begin{align} 
\begin{split} 
\tilde{\nu}_2 & =  \frac{2i}{\pi} \int_{-  k_\Delta }^{k_\Delta  } \, dk \,  \text{Tr} \, [P_B C(k)  P_A  \frac{d}{dk} C(k)] \simeq 0, \\
\end{split}
\label{eq:A1}
\end{align}
where the integration interval $[-k_\Delta, \, k_\Delta]$ is controlled by the temperature assuming the Fermi function $f(E_n(k)) = 1 /[\exp(\beta E_n(k)) + 1]$ to be $0$ or $1$ outside of this region for every band $n$. This interval can be made arbitrary small by lowering the temperature but it is always much larger than the momentum discretization scale $dk$. Note that $C(k) = f(H(k)) = [\exp(\beta H(k)) + 1]^{-1}$ is a well behaved function: If $H(k)$ is continuous and differentiable then $C(k)$ is also continuous and differentiable.

The correlation function can be decomposed as follows
\begin{align} 
\begin{split} 
C(k)  = \sum_n P_n(k) \, f(E_n(k)), \\
\end{split}
\label{eq:A2}
\end{align}
where $P_n(k) = |n, k\rangle\langle n, k|$ is a projector to the band $|n, k\rangle$ associated with energy level $E_n(k)$ (indexed in increasing order).   
 
In order to show that $\tilde{\nu}_2 \simeq 0$ we first divide the integral in Eq.~(\ref{eq:A1}) into two parts
\begin{align} 
\begin{split} 
i_1 = \int_{- k_\Delta }^0 \, dk \,  \text{Tr} \, [P_B C(k)  P_A  \frac{d}{dk} C(k)]; \\
i_2 = \int_{ 0 }^{k_\Delta } \, dk \,  \text{Tr} \, [P_B C(k)  P_A  \frac{d}{dk} C(k)]; \\
\end{split}
\label{eq:A3}
\end{align}
and consider each term separately. Note that each projector $P_n(k)$ changes continuously within the integration intervals of $i_1$ and $i_2$. Also, the projectors do not depend on the temperature and therefore we can always make the intervals small enough to consider $P_n(k)$ be independent of $k$ up to a negligible correction. 

Thus,
\begin{align} 
\begin{split} 
&i_1   \simeq  \int_{- k_\Delta }^0 \, dk \,   \sum_{n, m} [ f(E_n(k)) \frac{d}{dk} f(E_m(k)) ] \\
&\ \ \ \ \ \ \ \ \ \ \ \ \  \ \ \ \ \ \ \ \ \text{Tr} \,[ P_B P_n(k) P_A P_m(k)]\\
 &= \frac{1}{4}\int_{- k_\Delta }^0 \, dk \,   \sum_{n} [ f(E_n(k))  - f(-E_n(k)) ]\frac{d}{dk} f(E_n(k)) ; \\
&i_2   \simeq  \int_{0}^{k_\Delta } \, dk \,   \sum_{n, m} [ f(E_n(k)) \frac{d}{dk} f(E_m(k)) ] \\
&\ \ \ \ \ \ \ \ \ \ \ \ \  \ \ \ \ \ \ \ \ \text{Tr} \,[ P_B P_n(k) P_A P_m(k)]\\
 &= \frac{1}{4}\int_{0}^{k_\Delta } \, dk \,   \sum_{n} [ f(E_n(k))  - f(-E_n(k)) ]\frac{d}{dk} f(E_n(k))  , \\
\end{split}
\label{eq:A4}
\end{align}
where $\text{Tr} \,[ P_B P_n(k) P_A P_m(k)] = 1/4 \,  (\delta_{n, m} - \delta_{Sn, m})$ with $Sn$ denoting the complementary band to $n$, explicitly $P_{Sn} = S P_n S$. Here we also used orthonormality of the eigenstates $\langle n, k|m, k\rangle = \delta_{m, n}$ and  $\langle n, k|S|m, k\rangle = \delta_{m, Sn}$ along with identities $P_A + P_B = \mathbb{1}$, $P_A - P_B = S$, $P_n(k) = |n, k\rangle\langle n, k|$, and $E_{Sn}(k) = - E_{n}(k)$.

Finally, we notice that $i_1$ and $i_2$ in Eq.~(\ref{eq:A4}) do not depend on the projectors $P_n(k)$ and therefore:  
\begin{align} 
\begin{split} 
\tilde{\nu}_2 & = \frac{2i}{\pi} (i_1 + i_2) \\
& = \frac{i}{2\pi}  \sum_{n} \int_{- k_\Delta }^{k_\Delta } \, dk \,  [ f(E_n(k))  - f(-E_n(k)) ]\frac{d}{dk} f(E_n(k))  \\
& = \frac{i}{2\pi}  \sum_{n} \int_{f(E_n(-k_\Delta))}^{f(E_n(k_\Delta))} \, dx \,  [ 2x  - 1 ] \simeq 0,  \\
\end{split}
\label{eq:A5}
\end{align}
where we assume the temperature to be small enough so that $f(E_n(-k_\Delta)) \simeq  f(E_n(k_\Delta))$, and it was used that $f(-E_n(k)) = 1 - f(E_n(k))$. It follows that the contribution $\tilde{\nu}_2$ in Eq.~(\ref{eq:winding_k_parts2}) is indeed negligible in the zero-temperature limit.

\section{The generalized winding number represented within position space}\label{app:generalizedWinding}

Take a translationally invariant system described by a chiral Hamiltonian $H$ in 1D. The corresponding real-space correlation matrix $C$ can be presented as an inverse discrete Fourier transform of the correlation matrix $C(k)$ given within momentum space, thus,  
\begin{align} 
\begin{split} 
C_{j,j'} =  \frac{1}{L} \sum_{n=1}^{L} \exp(i  (j-j') k_n ) C(k_n) , \\
\end{split}
\label{eq:BFC1}
\end{align}
where $L$ is the total number of unit cells and $k_n = 2\pi n/L$.

We consider now the commutator between the position operator, $X = \sum_j j |j\rangle \langle j|$, and the correlation matrix $[X, C] = XC - CX$. This can be represented as follows
\begin{align} 
\begin{split} 
[X, C]_{j,j'} &=  \frac{1}{L} \sum_{n=1}^{L} (j-j^\prime) \exp(i (j-j') k_n) C(k_n) .
\end{split}
\label{eq:BFC2}
\end{align}
Considering elements in the bulk ($j-j'<<L$) and taking the thermodynamic limit we have
\begin{align} 
\begin{split} 
[X, C]_{j,j'} &=   -\frac{i}{2\pi}  \int_0^{2\pi} \, dk \, [ \frac{d}{dk}\exp(i k (j-j^\prime)  )] C(k), \\
&=   - \frac{i}{2\pi}  \int_0^{2\pi} \, dk \, \exp(i k (j-j^\prime)  ) \frac{d}{dk} C(k), \\
\end{split}
\label{eq:BFC3}
\end{align}
where we used the fact that
\begin{equation}
\int^{2\pi}_0\, dk \, {d} [\exp(i k (j- j^\prime) ) C(k)] / dk = 0
\end{equation}
because $C(k)$ is momentum-periodic. Note that the limit is well defined only for finite $(j-j^\prime)$, i.e. only for entries in the bulk.

It follows that the generalized winding number $\tilde{\nu}$, Eq.~(\ref{eq:nu2}), can be equivalently represented within real-space as
\begin{align} 
\begin{split} 
 & - \frac{4}{(M-N)} \text{Tr}_{(N, M)} \, \{P_B C P_A [X, C] \}  \\
&= \frac{4i}{(M-N) L^2} \sum^{M}_{j = N} \sum^L_{m=1} \sum^L_{n, n^\prime = 1} e^{i m(k_n - k_{n^\prime})} \,\\
& \ \ \ \ \ \ \ \ \ \ \text{Tr} \, [P_B  C(k_n) P_A [\frac{d}{dk} C(k)]|_{k = k_{n^\prime}}]    \\
&= \frac{4i}{L} \sum_{n} \text{Tr} \, [P_B  C(k_n) P_A [\frac{d}{dk} C(k)]|_{k = k_{n}}]    \\
 &=  \frac{2i}{\pi}\int^{\pi}_{-\pi} \, dk \, \text{Tr} \, [P_B C(k) P_A \frac{d}{dk} C(k)] = \tilde{\nu},    \\
\end{split}
\label{eq:BnuR}
\end{align}
where $ \text{Tr}_{(N,M)}[A]=\sum_{j=N}^{M} \text{Tr} \, A_{jj}/(N-M)$ is a volume trace over a partial region in the bulk, and the thermodynamic limit is implied in the commutator. The equation above is independent of $M$ and $N$ as long as they are in the bulk so it is enough to consider a single unit cell. For disordered systems, however, $M$ and $N$ must define a large region in the bulk for $\tilde{\nu}$ to represent well the effect of disorder.

\section{Criticality-preserving perturbations }
\label{app:disorderingperturbations}

Any two-band chiral system, $H = - \sigma_z H \sigma_z$ with chiral operator $\sigma_z$, can be represented as follows
\begin{align} 
\begin{split} 
H & = \sum_{i,j} h_{i, j} | A, i\rangle \langle B, j | + \text{H.c.}, \\
\end{split}
\label{eq:H}
\end{align}
where $h_{i,j}$ are in general complex, $A$ and $B$ are the two subblatices.  

For a chiral system to be gapless there must exist at least one zero energy solution, say $| \psi \rangle$. Note that we can always superpose two zero-energy states $| \psi \rangle$ and $\sigma_z | \psi \rangle$ to obtain a state that is completely localized on one of the two subblatices. Thus, without loss of generality we can assume $| \psi \rangle$ to be of the following form
\begin{align} 
\begin{split} 
| \psi \rangle = \sum_i \psi_{B, i} | B, i\rangle,  \\
\end{split}
\label{eq:psi}
\end{align}
where $\psi_{B, i}$ are some complex amplitudes that cannot be all zero.

To fulfil the zero-energy condition $H| \psi \rangle = 0$ the following constrains have to be satisfied:
\begin{align} 
\begin{split} 
 \sum_{j} h_{i, j} \psi_{B, j} = 0   \\
\end{split}
\label{eq:zerocond}
\end{align}
for all $i$. 

\textit{Case 1. $h_{i, i - 1} = h_{i, i}$ and $h_{i, j} = 0$ otherwise:} In this case the system is gapless because there is a solution to Eq.~(\ref{eq:zerocond}) given by $\psi_{B, i - 1} = - \psi_{B, i}$ for all $i$.

\textit{Case 2. $h_{i, i - 1} = h_{i, i} = -2 \, h_{i, i + 1}$ and $h_{i, j} = 0$ otherwise:} The system is gapless because $\psi_{B, i - 1} = \psi_{B, i}$ is a solution to Eq.~(\ref{eq:zerocond}).

In Sec.~IIE we consider these two cases and numerically verify stability of the generalized winding number $\tilde{\nu}$ to disordering perturbations that maintain the gap closed.

\section{Quantization of the generalized Zak phase at criticality}

To reach the result in Ref.~[\onlinecite{Bardyn2018}] for gapped systems a counting argument is required in order to see which elements of $M_T$ dominate the determinant. For gapless systems we have to do a similar counting argument, but one has to pay more attention around the gapless points. 

As mentioned in the main text we consider a single gapless point at $k = 0$, without loss of generality. The general eigenstates that fulfill chiral symmetry $S=\sigma_z$ have the form
\begin{align} 
\begin{split} 
| 1; {k} \rangle= \frac{1}{\sqrt{2}}
\begin{bmatrix}
e^{i \phi(k) }  \\
1 
\end{bmatrix}
\ \ \ \ \ 
| 2; {k} \rangle = \frac{1}{\sqrt{2}}
\begin{bmatrix}
-e^{i \phi(k) }  \\
1 
\end{bmatrix},
\end{split}
\end{align}
where the phase $\phi(k)$ is periodic and continuous everywhere except possibly at $k = 0$. The overlap elements are obtained as follows
\begin{align} 
\begin{split} 
\langle 1; k_j |1; k_{j + 1} \rangle &=  \frac{1}{2} [1 + \exp(i \phi_{j+1} - i\phi_{j} )] \\
\langle 1; k_j |2; k_{j + 1} \rangle &=  \frac{1}{2} [1 - \exp(i \phi_{j+1} - i \phi_{j} )]\\
\langle 2; k_j |1; k_{j + 1} \rangle &=  \frac{1}{2} [1 - \exp(i \phi_{j+1} - i\phi_{j} )]\\
\langle 2; k_j |2; k_{j + 1} \rangle &=  \frac{1}{2} [1 + \exp(-i \phi_{j+1} + i\phi_{j} )]\\
\end{split} 
\label{eq:overlaps}
\end{align}

We can now split $M_T$ into two terms and consider them separately. It follows
\begin{align} 
\begin{split} 
\text{Tr} \, M_T &= \text{Tr} \, \Pi_j \exp(-B_j) \, U_{j+1, j}\\
&= \text{Tr} [\, M_{0, L} \, M_{L - 1, 0} \,]\\
\end{split}
\label{eq:ZP4}
\end{align}
where
\begin{align} 
\begin{split}
M_{L-1, 0} &= \Pi_{j \in [0, L-1) } \, \exp(-\beta_{j+1} \sigma_z) \, U_{j+1, j}.\\
\end{split}
\label{eq:ZP5}
\end{align}
There are only two scenarios: $U_{0,L} = \mathbb{1}$, the bands are continuous across the gapless point, and (b) $U_{0,L} = \sigma_x$, the bands are flipped.

Let us consider first the continuous region,
\begin{align} 
\begin{split} 
 &M_{L-1, 0}\\=
&\begin{bmatrix}
e^{-\beta_{L-1}}\langle 1; k_{L-1} |...| 1; k_{0} \rangle & e^{-\beta_{L-1}}\langle 1; k_{L-1} |...| 2; k_{0} \rangle \\
e^{\beta_{L-1}}\langle 2; k_{L-1} |...| 1; k_{0}\rangle  & e^{\beta_{L-1}}\langle 2; k_{L-1} |...| 2; k_{0} \rangle
\end{bmatrix}, \\
\end{split}
\label{eq:case1}
\end{align}
where each of the terms also includes all possible ``jumps" between the two bands, with the appropriate Boltzmann factors. We first consider case (a) where the bands are continuous at the gapless point and $M_{0,L}=\mathbb{1}$. In this case it is clear that $W_\text{cond.} = \langle 2; k_0 | 2; k_{1} \rangle \langle 2; k_1 | ...| 2; k_{L-1} \rangle \langle 2; k_{L} | 2; k_{L} \rangle$ will dominate all other terms in the trace because it comes with the largest Boltzmann factors. The arguments for showing this are somewhat analogous to the explanation presented in Ref.~[\onlinecite{Bardyn2018}]. The low-temperature limit allows us to define some momentum scale $ k_\Delta \ll 2\pi$ around the gapless point outside of which the occupation of the bands can be approximated by $0$ and $1$ (in the region $k_\Delta< k <2\pi-k_\Delta$). In other words, $k_\Delta$ defines the neighborhood gapless interval where the occupations cannot be assumed to be precisely occupied or empty. This neighborhood region $[-k_\Delta, k_\Delta]$ will increase with $T$.

Take the thermodynamic limit for a finite temperature, $k_\Delta \gg \delta k$ and consider now $[M_{L-1,0}]_{22}$. The sum of all the terms with exactly two jumps between the bands, say jumps at $k_1$ and $k_2$, will be minor in comparison to the main term $W_\text{cond.}$: In total we have $L(L-1) \simeq (2 \pi / \delta k)^2$ such terms and each of them scales as $\sim \delta k^2$ so in general we can get a term of order $O(\delta k^0)$. However, from the definition of $k_\Delta$, the terms with  $k_1$ or $k_2$ outside of the neighborhood of the gapless point $-k_\Delta < k < k_\Delta$ will vanish, and the sum of remaining terms will be negligible because we are left with  $k_\Delta (2 k_\Delta - \delta k) / (\delta k)^2$ terms $\sim \delta k^2$, where $k_\Delta$ shrinks to zero as we take the limit of zero $T$.
It follows that the result in this case is analogous to the gapped case
\begin{align} 
\begin{split} 
\tilde{\phi}_\text{Zak} \simeq&  \left(\text{Im} \, \ln \, (-1)^{L+1}W_\text{cond.}\right) \\
\simeq \, & \phi_\text{Zak} + \pi(L+1).
\end{split}
\label{eq:Zak2A}
\end{align}

In case (b), where the bands flip and $U_{0,N}=\sigma_x$, we have
\begin{equation}
    \Tr M_T = [M_{L-1,0}]_{12}+[M_{L-1,0}]_{21}.
\end{equation}
These terms include at least one jump between bands 1 and 2. The terms with only one jump will dominate the trace for the same reasons as for the case with no band flipping. Take $[M_{L-1,0}]_{12}$, if the jump is far from the gapless point ($j_\text{jump} > \Delta$) it will result in an exponentially small term because of the Boltzmann factors. Similarly for the second term in the trace. In the neighborhood of the gapless region we must have one jump and therefore
\begin{align}
    \langle 1; k_{j+1} |2; k_{j} \rangle &=  \frac{1}{2} [1 - \exp(i \phi_{j} - i \phi_{j+1} )]\\
    &\simeq-\frac{1}{2}i\partial_k \phi(k)|_{k_j}\delta k,
\end{align}
while all other terms are of order $O(1)$. Therefore, the biggest contribution to the trace is given by

\begin{align} 
\begin{split} 
\text{Tr} \, M_T \simeq -i \delta_k B_{L -\Delta, \Delta} W^\text{cond.}_{\Delta<j<L -\Delta},
\end{split}
\label{eq:MT}
\end{align}
where 
\begin{align} 
\begin{split} 
W^\text{cond.}_{\Delta<j<L -\Delta}  = \langle 2; k_\Delta | 2; k_{\Delta + 1} \rangle \langle 2; k_{\Delta + 1} | ... | 2; k_{L - \Delta} \rangle
\end{split}
\label{eq:WDeltaA}
\end{align}
and 
\begin{align} 
\begin{split} 
B_{L -\Delta, \Delta} &= \partial_{k}\phi(k)|_{0+} \sum_{j \in (0, \Delta)} \, \exp(-\beta_{0, j} + \beta_{j, \Delta})  \\
&+ \partial_{k}\phi(k)|_{2\pi-} \sum_{j \in (L-\Delta, L)} \, \exp(\beta_{L-\Delta, j} - \beta_{j, L})\\
\end{split}
\label{eq:BN}
\end{align}
with $\beta_{j_1, j_2} = \sum_{j_1 < j < j_2} \, \beta_{j}$. Since  $\partial_{k}\phi(k)|_{0+} = \partial_{k}\phi(k)|_{2\pi-}$ . Thus, $B_{L -\Delta, \Delta}$ is a real constant. The low-temperature polarization in this case is therefore given by
\begin{align} 
\tilde{\phi}_\text{Zak} \simeq&  \left(\text{Im} \, \ln \, (-1)^{L+1}W^\text{cond.}_{\Delta<j<L -\Delta}\right) - \frac{\pi}{2}\text{sign}(B_{L -\Delta, \Delta}) \nonumber \\
\simeq&   \left(\text{Im} \, \ln \, (-1)^{L+1}W^\text{cond.}_{\Delta<j<L -\Delta}\right) - \frac{\pi}{2}\text{sign}(\partial_k \phi(k)|_{0^*}).
\label{eq:Zak3}
\end{align}
where $\Delta \ll L$. The first term is the Wilson loop obtained by excluding a neighborhood of the gapless region. This is equivalent to a winding number obtained in ~[\onlinecite{verresen2020}], which is quantized to half-integers for gapless phases. The second term complements this winding number making the polarization quantized to $0$ or $\pi$, similarly to gapped systems.

\bibliography{REF}

\begin{thebibliography}{39}%
\makeatletter
\providecommand \@ifxundefined [1]{%
 \@ifx{#1\undefined}
}%
\providecommand \@ifnum [1]{%
 \ifnum #1\expandafter \@firstoftwo
 \else \expandafter \@secondoftwo
 \fi
}%
\providecommand \@ifx [1]{%
 \ifx #1\expandafter \@firstoftwo
 \else \expandafter \@secondoftwo
 \fi
}%
\providecommand \natexlab [1]{#1}%
\providecommand \enquote  [1]{``#1''}%
\providecommand \bibnamefont  [1]{#1}%
\providecommand \bibfnamefont [1]{#1}%
\providecommand \citenamefont [1]{#1}%
\providecommand \href@noop [0]{\@secondoftwo}%
\providecommand \href [0]{\begingroup \@sanitize@url \@href}%
\providecommand \@href[1]{\@@startlink{#1}\@@href}%
\providecommand \@@href[1]{\endgroup#1\@@endlink}%
\providecommand \@sanitize@url [0]{\catcode `\\12\catcode `\$12\catcode
  `\&12\catcode `\#12\catcode `\^12\catcode `\_12\catcode `\%12\relax}%
\providecommand \@@startlink[1]{}%
\providecommand \@@endlink[0]{}%
\providecommand \url  [0]{\begingroup\@sanitize@url \@url }%
\providecommand \@url [1]{\endgroup\@href {#1}{\urlprefix }}%
\providecommand \urlprefix  [0]{URL }%
\providecommand \Eprint [0]{\href }%
\providecommand \doibase [0]{https://doi.org/}%
\providecommand \selectlanguage [0]{\@gobble}%
\providecommand \bibinfo  [0]{\@secondoftwo}%
\providecommand \bibfield  [0]{\@secondoftwo}%
\providecommand \translation [1]{[#1]}%
\providecommand \BibitemOpen [0]{}%
\providecommand \bibitemStop [0]{}%
\providecommand \bibitemNoStop [0]{.\EOS\space}%
\providecommand \EOS [0]{\spacefactor3000\relax}%
\providecommand \BibitemShut  [1]{\csname bibitem#1\endcsname}%
\let\auto@bib@innerbib\@empty
\bibitem [{\citenamefont {Kane}\ and\ \citenamefont
  {Mele}(2005)}]{kane_topological_2005}%
  \BibitemOpen
  \bibfield  {author} {\bibinfo {author} {\bibfnamefont {C.~L.}\ \bibnamefont
  {Kane}}\ and\ \bibinfo {author} {\bibfnamefont {E.~J.}\ \bibnamefont
  {Mele}},\ }\bibfield  {title} {\bibinfo {title} {{${Z}_{2}$ Topological Order
  and the Quantum Spin Hall Effect}},\ }\href
  {https://doi.org/10.1103/PhysRevLett.95.146802} {\bibfield  {journal}
  {\bibinfo  {journal} {Phys. Rev. Lett.}\ }\textbf {\bibinfo {volume} {95}},\
  \bibinfo {pages} {146802} (\bibinfo {year} {2005})}\BibitemShut {NoStop}%
\bibitem [{\citenamefont {Chiu}\ \emph {et~al.}(2016)\citenamefont {Chiu},
  \citenamefont {Teo}, \citenamefont {Schnyder},\ and\ \citenamefont
  {Ryu}}]{RuyReview}%
  \BibitemOpen
  \bibfield  {author} {\bibinfo {author} {\bibfnamefont {C.-K.}\ \bibnamefont
  {Chiu}}, \bibinfo {author} {\bibfnamefont {J.~C.~Y.}\ \bibnamefont {Teo}},
  \bibinfo {author} {\bibfnamefont {A.~P.}\ \bibnamefont {Schnyder}},\ and\
  \bibinfo {author} {\bibfnamefont {S.}~\bibnamefont {Ryu}},\ }\bibfield
  {title} {\bibinfo {title} {Classification of topological quantum matter with
  symmetries},\ }\href {https://doi.org/10.1103/RevModPhys.88.035005}
  {\bibfield  {journal} {\bibinfo  {journal} {Rev. Mod. Phys.}\ }\textbf
  {\bibinfo {volume} {88}},\ \bibinfo {pages} {035005} (\bibinfo {year}
  {2016})}\BibitemShut {NoStop}%
\bibitem [{\citenamefont {Altland}\ and\ \citenamefont
  {Zirnbauer}(1997)}]{AZ1997}%
  \BibitemOpen
  \bibfield  {author} {\bibinfo {author} {\bibfnamefont {A.}~\bibnamefont
  {Altland}}\ and\ \bibinfo {author} {\bibfnamefont {M.~R.}\ \bibnamefont
  {Zirnbauer}},\ }\bibfield  {title} {\bibinfo {title} {Nonstandard symmetry
  classes in mesoscopic normal-superconducting hybrid structures},\ }\href
  {https://doi.org/10.1103/PhysRevB.55.1142} {\bibfield  {journal} {\bibinfo
  {journal} {Phys. Rev. B}\ }\textbf {\bibinfo {volume} {55}},\ \bibinfo
  {pages} {1142} (\bibinfo {year} {1997})}\BibitemShut {NoStop}%
\bibitem [{\citenamefont {Schnyder}\ \emph {et~al.}(2008)\citenamefont
  {Schnyder}, \citenamefont {Ryu}, \citenamefont {Furusaki},\ and\
  \citenamefont {Ludwig}}]{schnyder_classification_2008}%
  \BibitemOpen
  \bibfield  {author} {\bibinfo {author} {\bibfnamefont {A.~P.}\ \bibnamefont
  {Schnyder}}, \bibinfo {author} {\bibfnamefont {S.}~\bibnamefont {Ryu}},
  \bibinfo {author} {\bibfnamefont {A.}~\bibnamefont {Furusaki}},\ and\
  \bibinfo {author} {\bibfnamefont {A.~W.~W.}\ \bibnamefont {Ludwig}},\
  }\bibfield  {title} {\bibinfo {title} {{Classification of topological
  insulators and superconductors in three spatial dimensions}},\ }\href
  {https://doi.org/10.1103/PhysRevB.78.195125} {\bibfield  {journal} {\bibinfo
  {journal} {Phys. Rev. B}\ }\textbf {\bibinfo {volume} {78}},\ \bibinfo
  {pages} {195125} (\bibinfo {year} {2008})}\BibitemShut {NoStop}%
\bibitem [{\citenamefont {Kitaev}(2009)}]{doi:10.1063/1.3149495}%
  \BibitemOpen
  \bibfield  {author} {\bibinfo {author} {\bibfnamefont {A.}~\bibnamefont
  {Kitaev}},\ }\bibfield  {title} {\bibinfo {title} {{Periodic table for
  topological insulators and superconductors}},\ }\href
  {https://doi.org/10.1063/1.3149495} {\bibfield  {journal} {\bibinfo
  {journal} {AIP Conference Proceedings}\ }\textbf {\bibinfo {volume} {1134}},\
  \bibinfo {pages} {22} (\bibinfo {year} {2009})},\ \Eprint
  {https://arxiv.org/abs/https://aip.scitation.org/doi/pdf/10.1063/1.3149495}
  {https://aip.scitation.org/doi/pdf/10.1063/1.3149495} \BibitemShut {NoStop}%
\bibitem [{\citenamefont {Resta}(1992)}]{Resta_polarization_1992}%
  \BibitemOpen
  \bibfield  {author} {\bibinfo {author} {\bibfnamefont {R.}~\bibnamefont
  {Resta}},\ }\bibfield  {title} {\bibinfo {title} {{Theory of the electric
  polarization in crystals}},\ }\href
  {https://doi.org/10.1080/00150199208016065} {\bibfield  {journal} {\bibinfo
  {journal} {Ferroelectrics}\ }\textbf {\bibinfo {volume} {136}},\ \bibinfo
  {pages} {51} (\bibinfo {year} {1992})},\ \Eprint
  {https://arxiv.org/abs/https://doi.org/10.1080/00150199208016065}
  {https://doi.org/10.1080/00150199208016065} \BibitemShut {NoStop}%
\bibitem [{\citenamefont {King-Smith}\ and\ \citenamefont
  {Vanderbilt}(1993)}]{KingSmith_polarization_1993}%
  \BibitemOpen
  \bibfield  {author} {\bibinfo {author} {\bibfnamefont {R.~D.}\ \bibnamefont
  {King-Smith}}\ and\ \bibinfo {author} {\bibfnamefont {D.}~\bibnamefont
  {Vanderbilt}},\ }\bibfield  {title} {\bibinfo {title} {{Theory of
  polarization of crystalline solids}},\ }\href
  {https://doi.org/10.1103/PhysRevB.47.1651} {\bibfield  {journal} {\bibinfo
  {journal} {Phys. Rev. B}\ }\textbf {\bibinfo {volume} {47}},\ \bibinfo
  {pages} {1651} (\bibinfo {year} {1993})}\BibitemShut {NoStop}%
\bibitem [{\citenamefont {Vanderbilt}\ and\ \citenamefont
  {King-Smith}(1993)}]{Vanderbilt_polarization_1993}%
  \BibitemOpen
  \bibfield  {author} {\bibinfo {author} {\bibfnamefont {D.}~\bibnamefont
  {Vanderbilt}}\ and\ \bibinfo {author} {\bibfnamefont {R.~D.}\ \bibnamefont
  {King-Smith}},\ }\bibfield  {title} {\bibinfo {title} {{Electric polarization
  as a bulk quantity and its relation to surface charge}},\ }\href
  {https://doi.org/10.1103/PhysRevB.48.4442} {\bibfield  {journal} {\bibinfo
  {journal} {Phys. Rev. B}\ }\textbf {\bibinfo {volume} {48}},\ \bibinfo
  {pages} {4442} (\bibinfo {year} {1993})}\BibitemShut {NoStop}%
\bibitem [{\citenamefont {Resta}(1998)}]{Resta}%
  \BibitemOpen
  \bibfield  {author} {\bibinfo {author} {\bibfnamefont {R.}~\bibnamefont
  {Resta}},\ }\bibfield  {title} {\bibinfo {title} {Quantum-mechanical position
  operator in extended systems},\ }\href
  {https://doi.org/10.1103/PhysRevLett.80.1800} {\bibfield  {journal} {\bibinfo
   {journal} {Phys. Rev. Lett.}\ }\textbf {\bibinfo {volume} {80}},\ \bibinfo
  {pages} {1800} (\bibinfo {year} {1998})}\BibitemShut {NoStop}%
\bibitem [{\citenamefont {Watanabe}\ and\ \citenamefont
  {Oshikawa}(2018)}]{Watanabe_polarization_2018}%
  \BibitemOpen
  \bibfield  {author} {\bibinfo {author} {\bibfnamefont {H.}~\bibnamefont
  {Watanabe}}\ and\ \bibinfo {author} {\bibfnamefont {M.}~\bibnamefont
  {Oshikawa}},\ }\bibfield  {title} {\bibinfo {title} {{Inequivalent Berry
  Phases for the Bulk Polarization}},\ }\href
  {https://doi.org/10.1103/PhysRevX.8.021065} {\bibfield  {journal} {\bibinfo
  {journal} {Phys. Rev. X}\ }\textbf {\bibinfo {volume} {8}},\ \bibinfo {pages}
  {021065} (\bibinfo {year} {2018})}\BibitemShut {NoStop}%
\bibitem [{\citenamefont {Kestner}\ \emph {et~al.}(2011)\citenamefont
  {Kestner}, \citenamefont {Wang}, \citenamefont {Sau},\ and\ \citenamefont
  {Das~Sarma}}]{Kestner}%
  \BibitemOpen
  \bibfield  {author} {\bibinfo {author} {\bibfnamefont {J.~P.}\ \bibnamefont
  {Kestner}}, \bibinfo {author} {\bibfnamefont {B.}~\bibnamefont {Wang}},
  \bibinfo {author} {\bibfnamefont {J.~D.}\ \bibnamefont {Sau}},\ and\ \bibinfo
  {author} {\bibfnamefont {S.}~\bibnamefont {Das~Sarma}},\ }\bibfield  {title}
  {\bibinfo {title} {Prediction of a gapless topological haldane liquid phase
  in a one-dimensional cold polar molecular lattice},\ }\href
  {https://doi.org/10.1103/PhysRevB.83.174409} {\bibfield  {journal} {\bibinfo
  {journal} {Phys. Rev. B}\ }\textbf {\bibinfo {volume} {83}},\ \bibinfo
  {pages} {174409} (\bibinfo {year} {2011})}\BibitemShut {NoStop}%
\bibitem [{\citenamefont {Cheng}\ and\ \citenamefont {Tu}(2011)}]{Cheng}%
  \BibitemOpen
  \bibfield  {author} {\bibinfo {author} {\bibfnamefont {M.}~\bibnamefont
  {Cheng}}\ and\ \bibinfo {author} {\bibfnamefont {H.-H.}\ \bibnamefont {Tu}},\
  }\bibfield  {title} {\bibinfo {title} {Majorana edge states in interacting
  two-chain ladders of fermions},\ }\href
  {https://doi.org/10.1103/PhysRevB.84.094503} {\bibfield  {journal} {\bibinfo
  {journal} {Phys. Rev. B}\ }\textbf {\bibinfo {volume} {84}},\ \bibinfo
  {pages} {094503} (\bibinfo {year} {2011})}\BibitemShut {NoStop}%
\bibitem [{\citenamefont {Fidkowski}\ \emph {et~al.}(2011)\citenamefont
  {Fidkowski}, \citenamefont {Lutchyn}, \citenamefont {Nayak},\ and\
  \citenamefont {Fisher}}]{Fidkowski}%
  \BibitemOpen
  \bibfield  {author} {\bibinfo {author} {\bibfnamefont {L.}~\bibnamefont
  {Fidkowski}}, \bibinfo {author} {\bibfnamefont {R.~M.}\ \bibnamefont
  {Lutchyn}}, \bibinfo {author} {\bibfnamefont {C.}~\bibnamefont {Nayak}},\
  and\ \bibinfo {author} {\bibfnamefont {M.~P.~A.}\ \bibnamefont {Fisher}},\
  }\bibfield  {title} {\bibinfo {title} {Majorana zero modes in one-dimensional
  quantum wires without long-ranged superconducting order},\ }\href
  {https://doi.org/10.1103/PhysRevB.84.195436} {\bibfield  {journal} {\bibinfo
  {journal} {Phys. Rev. B}\ }\textbf {\bibinfo {volume} {84}},\ \bibinfo
  {pages} {195436} (\bibinfo {year} {2011})}\BibitemShut {NoStop}%
\bibitem [{\citenamefont {Sau}\ \emph {et~al.}(2011)\citenamefont {Sau},
  \citenamefont {Halperin}, \citenamefont {Flensberg},\ and\ \citenamefont
  {Das~Sarma}}]{Sau}%
  \BibitemOpen
  \bibfield  {author} {\bibinfo {author} {\bibfnamefont {J.~D.}\ \bibnamefont
  {Sau}}, \bibinfo {author} {\bibfnamefont {B.~I.}\ \bibnamefont {Halperin}},
  \bibinfo {author} {\bibfnamefont {K.}~\bibnamefont {Flensberg}},\ and\
  \bibinfo {author} {\bibfnamefont {S.}~\bibnamefont {Das~Sarma}},\ }\bibfield
  {title} {\bibinfo {title} {Number conserving theory for topologically
  protected degeneracy in one-dimensional fermions},\ }\href
  {https://doi.org/10.1103/PhysRevB.84.144509} {\bibfield  {journal} {\bibinfo
  {journal} {Phys. Rev. B}\ }\textbf {\bibinfo {volume} {84}},\ \bibinfo
  {pages} {144509} (\bibinfo {year} {2011})}\BibitemShut {NoStop}%
\bibitem [{\citenamefont {Kraus}\ \emph {et~al.}(2013)\citenamefont {Kraus},
  \citenamefont {Dalmonte}, \citenamefont {Baranov}, \citenamefont
  {L\"auchli},\ and\ \citenamefont {Zoller}}]{Kraus}%
  \BibitemOpen
  \bibfield  {author} {\bibinfo {author} {\bibfnamefont {C.~V.}\ \bibnamefont
  {Kraus}}, \bibinfo {author} {\bibfnamefont {M.}~\bibnamefont {Dalmonte}},
  \bibinfo {author} {\bibfnamefont {M.~A.}\ \bibnamefont {Baranov}}, \bibinfo
  {author} {\bibfnamefont {A.~M.}\ \bibnamefont {L\"auchli}},\ and\ \bibinfo
  {author} {\bibfnamefont {P.}~\bibnamefont {Zoller}},\ }\bibfield  {title}
  {\bibinfo {title} {Majorana edge states in atomic wires coupled by pair
  hopping},\ }\href {https://doi.org/10.1103/PhysRevLett.111.173004} {\bibfield
   {journal} {\bibinfo  {journal} {Phys. Rev. Lett.}\ }\textbf {\bibinfo
  {volume} {111}},\ \bibinfo {pages} {173004} (\bibinfo {year}
  {2013})}\BibitemShut {NoStop}%
\bibitem [{\citenamefont {Keselman}\ and\ \citenamefont
  {Berg}(2015)}]{Keselman1}%
  \BibitemOpen
  \bibfield  {author} {\bibinfo {author} {\bibfnamefont {A.}~\bibnamefont
  {Keselman}}\ and\ \bibinfo {author} {\bibfnamefont {E.}~\bibnamefont
  {Berg}},\ }\bibfield  {title} {\bibinfo {title} {Gapless symmetry-protected
  topological phase of fermions in one dimension},\ }\href
  {https://doi.org/10.1103/PhysRevB.91.235309} {\bibfield  {journal} {\bibinfo
  {journal} {Phys. Rev. B}\ }\textbf {\bibinfo {volume} {91}},\ \bibinfo
  {pages} {235309} (\bibinfo {year} {2015})}\BibitemShut {NoStop}%
\bibitem [{\citenamefont {Iemini}\ \emph {et~al.}(2015)\citenamefont {Iemini},
  \citenamefont {Mazza}, \citenamefont {Rossini}, \citenamefont {Fazio},\ and\
  \citenamefont {Diehl}}]{Iemini}%
  \BibitemOpen
  \bibfield  {author} {\bibinfo {author} {\bibfnamefont {F.}~\bibnamefont
  {Iemini}}, \bibinfo {author} {\bibfnamefont {L.}~\bibnamefont {Mazza}},
  \bibinfo {author} {\bibfnamefont {D.}~\bibnamefont {Rossini}}, \bibinfo
  {author} {\bibfnamefont {R.}~\bibnamefont {Fazio}},\ and\ \bibinfo {author}
  {\bibfnamefont {S.}~\bibnamefont {Diehl}},\ }\bibfield  {title} {\bibinfo
  {title} {Localized majorana-like modes in a number-conserving setting: An
  exactly solvable model},\ }\href
  {https://doi.org/10.1103/PhysRevLett.115.156402} {\bibfield  {journal}
  {\bibinfo  {journal} {Phys. Rev. Lett.}\ }\textbf {\bibinfo {volume} {115}},\
  \bibinfo {pages} {156402} (\bibinfo {year} {2015})}\BibitemShut {NoStop}%
\bibitem [{\citenamefont {Lang}\ and\ \citenamefont {B\"uchler}(2015)}]{Lang}%
  \BibitemOpen
  \bibfield  {author} {\bibinfo {author} {\bibfnamefont {N.}~\bibnamefont
  {Lang}}\ and\ \bibinfo {author} {\bibfnamefont {H.~P.}\ \bibnamefont
  {B\"uchler}},\ }\bibfield  {title} {\bibinfo {title} {Topological states in a
  microscopic model of interacting fermions},\ }\href
  {https://doi.org/10.1103/PhysRevB.92.041118} {\bibfield  {journal} {\bibinfo
  {journal} {Phys. Rev. B}\ }\textbf {\bibinfo {volume} {92}},\ \bibinfo
  {pages} {041118} (\bibinfo {year} {2015})}\BibitemShut {NoStop}%
\bibitem [{\citenamefont {Montorsi}\ \emph {et~al.}(2017)\citenamefont
  {Montorsi}, \citenamefont {Dolcini}, \citenamefont {Iotti},\ and\
  \citenamefont {Rossi}}]{Montorsi}%
  \BibitemOpen
  \bibfield  {author} {\bibinfo {author} {\bibfnamefont {A.}~\bibnamefont
  {Montorsi}}, \bibinfo {author} {\bibfnamefont {F.}~\bibnamefont {Dolcini}},
  \bibinfo {author} {\bibfnamefont {R.~C.}\ \bibnamefont {Iotti}},\ and\
  \bibinfo {author} {\bibfnamefont {F.}~\bibnamefont {Rossi}},\ }\bibfield
  {title} {\bibinfo {title} {Symmetry-protected topological phases of
  one-dimensional interacting fermions with spin-charge separation},\ }\href
  {https://doi.org/10.1103/PhysRevB.95.245108} {\bibfield  {journal} {\bibinfo
  {journal} {Phys. Rev. B}\ }\textbf {\bibinfo {volume} {95}},\ \bibinfo
  {pages} {245108} (\bibinfo {year} {2017})}\BibitemShut {NoStop}%
\bibitem [{\citenamefont {Ruhman}\ and\ \citenamefont {Altman}(2017)}]{Ruhman}%
  \BibitemOpen
  \bibfield  {author} {\bibinfo {author} {\bibfnamefont {J.}~\bibnamefont
  {Ruhman}}\ and\ \bibinfo {author} {\bibfnamefont {E.}~\bibnamefont
  {Altman}},\ }\bibfield  {title} {\bibinfo {title} {Topological degeneracy and
  pairing in a one-dimensional gas of spinless fermions},\ }\href
  {https://doi.org/10.1103/PhysRevB.96.085133} {\bibfield  {journal} {\bibinfo
  {journal} {Phys. Rev. B}\ }\textbf {\bibinfo {volume} {96}},\ \bibinfo
  {pages} {085133} (\bibinfo {year} {2017})}\BibitemShut {NoStop}%
\bibitem [{\citenamefont {Jiang}\ \emph {et~al.}(2018)\citenamefont {Jiang},
  \citenamefont {Li}, \citenamefont {Seidel},\ and\ \citenamefont
  {Lee}}]{Jiang}%
  \BibitemOpen
  \bibfield  {author} {\bibinfo {author} {\bibfnamefont {H.-C.}\ \bibnamefont
  {Jiang}}, \bibinfo {author} {\bibfnamefont {Z.-X.}\ \bibnamefont {Li}},
  \bibinfo {author} {\bibfnamefont {A.}~\bibnamefont {Seidel}},\ and\ \bibinfo
  {author} {\bibfnamefont {D.-H.}\ \bibnamefont {Lee}},\ }\bibfield  {title}
  {\bibinfo {title} {Symmetry protected topological luttinger liquids and the
  phase transition between them},\ }\href
  {https://doi.org/https://doi.org/10.1016/j.scib.2018.05.010} {\bibfield
  {journal} {\bibinfo  {journal} {Science Bulletin}\ }\textbf {\bibinfo
  {volume} {63}},\ \bibinfo {pages} {753} (\bibinfo {year} {2018})}\BibitemShut
  {NoStop}%
\bibitem [{\citenamefont {Zhang}\ and\ \citenamefont {Liu}(2018)}]{Zhang}%
  \BibitemOpen
  \bibfield  {author} {\bibinfo {author} {\bibfnamefont {R.-X.}\ \bibnamefont
  {Zhang}}\ and\ \bibinfo {author} {\bibfnamefont {C.-X.}\ \bibnamefont
  {Liu}},\ }\bibfield  {title} {\bibinfo {title} {Crystalline
  symmetry-protected majorana mode in number-conserving dirac semimetal
  nanowires},\ }\href {https://doi.org/10.1103/PhysRevLett.120.156802}
  {\bibfield  {journal} {\bibinfo  {journal} {Phys. Rev. Lett.}\ }\textbf
  {\bibinfo {volume} {120}},\ \bibinfo {pages} {156802} (\bibinfo {year}
  {2018})}\BibitemShut {NoStop}%
\bibitem [{\citenamefont {Parker}\ \emph {et~al.}(2018)\citenamefont {Parker},
  \citenamefont {Scaffidi},\ and\ \citenamefont {Vasseur}}]{Parker}%
  \BibitemOpen
  \bibfield  {author} {\bibinfo {author} {\bibfnamefont {D.~E.}\ \bibnamefont
  {Parker}}, \bibinfo {author} {\bibfnamefont {T.}~\bibnamefont {Scaffidi}},\
  and\ \bibinfo {author} {\bibfnamefont {R.}~\bibnamefont {Vasseur}},\
  }\bibfield  {title} {\bibinfo {title} {Topological luttinger liquids from
  decorated domain walls},\ }\href {https://doi.org/10.1103/PhysRevB.97.165114}
  {\bibfield  {journal} {\bibinfo  {journal} {Phys. Rev. B}\ }\textbf {\bibinfo
  {volume} {97}},\ \bibinfo {pages} {165114} (\bibinfo {year}
  {2018})}\BibitemShut {NoStop}%
\bibitem [{\citenamefont {Keselman}\ \emph {et~al.}(2018)\citenamefont
  {Keselman}, \citenamefont {Berg},\ and\ \citenamefont {Azaria}}]{Keselman2}%
  \BibitemOpen
  \bibfield  {author} {\bibinfo {author} {\bibfnamefont {A.}~\bibnamefont
  {Keselman}}, \bibinfo {author} {\bibfnamefont {E.}~\bibnamefont {Berg}},\
  and\ \bibinfo {author} {\bibfnamefont {P.}~\bibnamefont {Azaria}},\
  }\bibfield  {title} {\bibinfo {title} {From one-dimensional charge conserving
  superconductors to the gapless haldane phase},\ }\href
  {https://doi.org/10.1103/PhysRevB.98.214501} {\bibfield  {journal} {\bibinfo
  {journal} {Phys. Rev. B}\ }\textbf {\bibinfo {volume} {98}},\ \bibinfo
  {pages} {214501} (\bibinfo {year} {2018})}\BibitemShut {NoStop}%
\bibitem [{\citenamefont {Verresen}\ \emph {et~al.}(2018)\citenamefont
  {Verresen}, \citenamefont {Jones},\ and\ \citenamefont
  {Pollmann}}]{verresen2018}%
  \BibitemOpen
  \bibfield  {author} {\bibinfo {author} {\bibfnamefont {R.}~\bibnamefont
  {Verresen}}, \bibinfo {author} {\bibfnamefont {N.~G.}\ \bibnamefont
  {Jones}},\ and\ \bibinfo {author} {\bibfnamefont {F.}~\bibnamefont
  {Pollmann}},\ }\bibfield  {title} {\bibinfo {title} {Topology and edge modes
  in quantum critical chains},\ }\href
  {https://doi.org/10.1103/PhysRevLett.120.057001} {\bibfield  {journal}
  {\bibinfo  {journal} {Phys. Rev. Lett.}\ }\textbf {\bibinfo {volume} {120}},\
  \bibinfo {pages} {057001} (\bibinfo {year} {2018})}\BibitemShut {NoStop}%
\bibitem [{\citenamefont {Balabanov}\ \emph {et~al.}(2021)\citenamefont
  {Balabanov}, \citenamefont {Erkensten},\ and\ \citenamefont
  {Johannesson}}]{BEH}%
  \BibitemOpen
  \bibfield  {author} {\bibinfo {author} {\bibfnamefont {O.}~\bibnamefont
  {Balabanov}}, \bibinfo {author} {\bibfnamefont {D.}~\bibnamefont
  {Erkensten}},\ and\ \bibinfo {author} {\bibfnamefont {H.}~\bibnamefont
  {Johannesson}},\ }\bibfield  {title} {\bibinfo {title} {Topology of critical
  chiral phases: Multiband insulators and superconductors},\ }\href
  {https://doi.org/10.1103/PhysRevResearch.3.043048} {\bibfield  {journal}
  {\bibinfo  {journal} {Phys. Rev. Research}\ }\textbf {\bibinfo {volume}
  {3}},\ \bibinfo {pages} {043048} (\bibinfo {year} {2021})}\BibitemShut
  {NoStop}%
\bibitem [{\citenamefont {Verresen}\ \emph {et~al.}(2021)\citenamefont
  {Verresen}, \citenamefont {Thorngren}, \citenamefont {Jones},\ and\
  \citenamefont {Pollmann}}]{verresenPRX}%
  \BibitemOpen
  \bibfield  {author} {\bibinfo {author} {\bibfnamefont {R.}~\bibnamefont
  {Verresen}}, \bibinfo {author} {\bibfnamefont {R.}~\bibnamefont {Thorngren}},
  \bibinfo {author} {\bibfnamefont {N.~G.}\ \bibnamefont {Jones}},\ and\
  \bibinfo {author} {\bibfnamefont {F.}~\bibnamefont {Pollmann}},\ }\bibfield
  {title} {\bibinfo {title} {Gapless topological phases and symmetry-enriched
  quantum criticality},\ }\href {https://doi.org/10.1103/PhysRevX.11.041059}
  {\bibfield  {journal} {\bibinfo  {journal} {Phys. Rev. X}\ }\textbf {\bibinfo
  {volume} {11}},\ \bibinfo {pages} {041059} (\bibinfo {year}
  {2021})}\BibitemShut {NoStop}%
\bibitem [{\citenamefont {Verresen}(2020)}]{verresen2020}%
  \BibitemOpen
  \bibfield  {author} {\bibinfo {author} {\bibfnamefont {R.}~\bibnamefont
  {Verresen}},\ }\href@noop {} {\bibinfo {title} {Topology and edge states
  survive quantum criticality between topological insulators}} (\bibinfo {year}
  {2020}),\ \Eprint {https://arxiv.org/abs/2003.05453} {arXiv:2003.05453
  [cond-mat.str-el]} \BibitemShut {NoStop}%
\bibitem [{\citenamefont {Jones}\ and\ \citenamefont {Verresen}(2019)}]{Jones}%
  \BibitemOpen
  \bibfield  {author} {\bibinfo {author} {\bibfnamefont {N.}~\bibnamefont
  {Jones}}\ and\ \bibinfo {author} {\bibfnamefont {R.}~\bibnamefont
  {Verresen}},\ }\bibfield  {title} {\bibinfo {title} {Asymptotic correlations
  in gapped and critical topological phases of 1d quantum systems},\ }\href
  {https://doi.org/10.1007/s10955-019-02257-9} {\bibfield  {journal} {\bibinfo
  {journal} {Journal of Statistical Physics}\ }\textbf {\bibinfo {volume}
  {175}} (\bibinfo {year} {2019})}\BibitemShut {NoStop}%
\bibitem [{\citenamefont {Duque}\ \emph {et~al.}(2021)\citenamefont {Duque},
  \citenamefont {Hu}, \citenamefont {You}, \citenamefont {Khemani},
  \citenamefont {Verresen},\ and\ \citenamefont {Vasseur}}]{Duque}%
  \BibitemOpen
  \bibfield  {author} {\bibinfo {author} {\bibfnamefont {C.~M.}\ \bibnamefont
  {Duque}}, \bibinfo {author} {\bibfnamefont {H.-Y.}\ \bibnamefont {Hu}},
  \bibinfo {author} {\bibfnamefont {Y.-Z.}\ \bibnamefont {You}}, \bibinfo
  {author} {\bibfnamefont {V.}~\bibnamefont {Khemani}}, \bibinfo {author}
  {\bibfnamefont {R.}~\bibnamefont {Verresen}},\ and\ \bibinfo {author}
  {\bibfnamefont {R.}~\bibnamefont {Vasseur}},\ }\bibfield  {title} {\bibinfo
  {title} {Topological and symmetry-enriched random quantum critical points},\
  }\href {https://doi.org/10.1103/PhysRevB.103.L100207} {\bibfield  {journal}
  {\bibinfo  {journal} {Phys. Rev. B}\ }\textbf {\bibinfo {volume} {103}},\
  \bibinfo {pages} {L100207} (\bibinfo {year} {2021})}\BibitemShut {NoStop}%
\bibitem [{\citenamefont {Kumar}\ \emph
  {et~al.}(2021{\natexlab{a}})\citenamefont {Kumar}, \citenamefont {Kartik},
  \citenamefont {Rahul},\ and\ \citenamefont {Sarkar}}]{Kumar1}%
  \BibitemOpen
  \bibfield  {author} {\bibinfo {author} {\bibfnamefont {R.~R.}\ \bibnamefont
  {Kumar}}, \bibinfo {author} {\bibfnamefont {Y.~R.}\ \bibnamefont {Kartik}},
  \bibinfo {author} {\bibfnamefont {S.}~\bibnamefont {Rahul}},\ and\ \bibinfo
  {author} {\bibfnamefont {S.}~\bibnamefont {Sarkar}},\ }\bibfield  {title}
  {\bibinfo {title} {Multi-critical topological transition at quantum
  criticality},\ }\bibfield  {journal} {\bibinfo  {journal} {Scientific
  Reports}\ }\textbf {\bibinfo {volume} {11}},\ \href
  {https://doi.org/10.1038/s41598-020-80337-7} {10.1038/s41598-020-80337-7}
  (\bibinfo {year} {2021}{\natexlab{a}})\BibitemShut {NoStop}%
\bibitem [{\citenamefont {Kumar}\ \emph
  {et~al.}(2021{\natexlab{b}})\citenamefont {Kumar}, \citenamefont {Roy},
  \citenamefont {Kartik}, \citenamefont {Rahul},\ and\ \citenamefont
  {Sarkar}}]{Kumar2}%
  \BibitemOpen
  \bibfield  {author} {\bibinfo {author} {\bibfnamefont {R.~R.}\ \bibnamefont
  {Kumar}}, \bibinfo {author} {\bibfnamefont {N.}~\bibnamefont {Roy}}, \bibinfo
  {author} {\bibfnamefont {Y.~R.}\ \bibnamefont {Kartik}}, \bibinfo {author}
  {\bibfnamefont {S.}~\bibnamefont {Rahul}},\ and\ \bibinfo {author}
  {\bibfnamefont {S.}~\bibnamefont {Sarkar}},\ }\href@noop {} {\bibinfo {title}
  {Topological phase transition at quantum criticality}} (\bibinfo {year}
  {2021}{\natexlab{b}}),\ \Eprint {https://arxiv.org/abs/2112.02485}
  {arXiv:2112.02485 [cond-mat.str-el]} \BibitemShut {NoStop}%
\bibitem [{\citenamefont {Tantivasadakarn}\ \emph {et~al.}(2021)\citenamefont
  {Tantivasadakarn}, \citenamefont {Thorngren}, \citenamefont {Vishwanath},\
  and\ \citenamefont {Verresen}}]{Tantivasadakarn}%
  \BibitemOpen
  \bibfield  {author} {\bibinfo {author} {\bibfnamefont {N.}~\bibnamefont
  {Tantivasadakarn}}, \bibinfo {author} {\bibfnamefont {R.}~\bibnamefont
  {Thorngren}}, \bibinfo {author} {\bibfnamefont {A.}~\bibnamefont
  {Vishwanath}},\ and\ \bibinfo {author} {\bibfnamefont {R.}~\bibnamefont
  {Verresen}},\ }\href@noop {} {\bibinfo {title} {Building models of
  topological quantum criticality from pivot hamiltonians}} (\bibinfo {year}
  {2021}),\ \Eprint {https://arxiv.org/abs/2110.09512} {arXiv:2110.09512
  [cond-mat.str-el]} \BibitemShut {NoStop}%
\bibitem [{\citenamefont {Ye}\ \emph {et~al.}(2021)\citenamefont {Ye},
  \citenamefont {Guo}, \citenamefont {He}, \citenamefont {Wang},\ and\
  \citenamefont {Zou}}]{Zou2021}%
  \BibitemOpen
  \bibfield  {author} {\bibinfo {author} {\bibfnamefont {W.}~\bibnamefont
  {Ye}}, \bibinfo {author} {\bibfnamefont {M.}~\bibnamefont {Guo}}, \bibinfo
  {author} {\bibfnamefont {Y.-C.}\ \bibnamefont {He}}, \bibinfo {author}
  {\bibfnamefont {C.}~\bibnamefont {Wang}},\ and\ \bibinfo {author}
  {\bibfnamefont {L.}~\bibnamefont {Zou}},\ }\href@noop {} {\bibinfo {title}
  {Topological characterization of lieb-schultz-mattis constraints and
  applications to symmetry-enriched quantum criticality}} (\bibinfo {year}
  {2021}),\ \Eprint {https://arxiv.org/abs/2111.12097} {arXiv:2111.12097
  [cond-mat.str-el]} \BibitemShut {NoStop}%
\bibitem [{\citenamefont {Bardyn}\ \emph {et~al.}(2018)\citenamefont {Bardyn},
  \citenamefont {Wawer}, \citenamefont {Altland}, \citenamefont
  {Fleischhauer},\ and\ \citenamefont {Diehl}}]{Bardyn2018}%
  \BibitemOpen
  \bibfield  {author} {\bibinfo {author} {\bibfnamefont {C.-E.}\ \bibnamefont
  {Bardyn}}, \bibinfo {author} {\bibfnamefont {L.}~\bibnamefont {Wawer}},
  \bibinfo {author} {\bibfnamefont {A.}~\bibnamefont {Altland}}, \bibinfo
  {author} {\bibfnamefont {M.}~\bibnamefont {Fleischhauer}},\ and\ \bibinfo
  {author} {\bibfnamefont {S.}~\bibnamefont {Diehl}},\ }\bibfield  {title}
  {\bibinfo {title} {Probing the topology of density matrices},\ }\href
  {https://doi.org/10.1103/PhysRevX.8.011035} {\bibfield  {journal} {\bibinfo
  {journal} {Phys. Rev. X}\ }\textbf {\bibinfo {volume} {8}},\ \bibinfo {pages}
  {011035} (\bibinfo {year} {2018})}\BibitemShut {NoStop}%
\bibitem [{\citenamefont {Mondragon-Shem}\ \emph {et~al.}(2014)\citenamefont
  {Mondragon-Shem}, \citenamefont {Hughes}, \citenamefont {Song},\ and\
  \citenamefont {Prodan}}]{real_nu1}%
  \BibitemOpen
  \bibfield  {author} {\bibinfo {author} {\bibfnamefont {I.}~\bibnamefont
  {Mondragon-Shem}}, \bibinfo {author} {\bibfnamefont {T.~L.}\ \bibnamefont
  {Hughes}}, \bibinfo {author} {\bibfnamefont {J.}~\bibnamefont {Song}},\ and\
  \bibinfo {author} {\bibfnamefont {E.}~\bibnamefont {Prodan}},\ }\bibfield
  {title} {\bibinfo {title} {Topological criticality in the chiral-symmetric
  aiii class at strong disorder},\ }\href
  {https://doi.org/10.1103/PhysRevLett.113.046802} {\bibfield  {journal}
  {\bibinfo  {journal} {Phys. Rev. Lett.}\ }\textbf {\bibinfo {volume} {113}},\
  \bibinfo {pages} {046802} (\bibinfo {year} {2014})}\BibitemShut {NoStop}%
\bibitem [{\citenamefont {Song}\ and\ \citenamefont {Prodan}(2014)}]{real_nu2}%
  \BibitemOpen
  \bibfield  {author} {\bibinfo {author} {\bibfnamefont {J.}~\bibnamefont
  {Song}}\ and\ \bibinfo {author} {\bibfnamefont {E.}~\bibnamefont {Prodan}},\
  }\bibfield  {title} {\bibinfo {title} {Aiii and bdi topological systems at
  strong disorder},\ }\href {https://doi.org/10.1103/PhysRevB.89.224203}
  {\bibfield  {journal} {\bibinfo  {journal} {Phys. Rev. B}\ }\textbf {\bibinfo
  {volume} {89}},\ \bibinfo {pages} {224203} (\bibinfo {year}
  {2014})}\BibitemShut {NoStop}%
\bibitem [{\citenamefont {Ortega-Taberner}\ and\ \citenamefont
  {Hermanns}(2021)}]{COTMH}%
  \BibitemOpen
  \bibfield  {author} {\bibinfo {author} {\bibfnamefont {C.}~\bibnamefont
  {Ortega-Taberner}}\ and\ \bibinfo {author} {\bibfnamefont {M.}~\bibnamefont
  {Hermanns}},\ }\bibfield  {title} {\bibinfo {title} {Relation of the
  entanglement spectrum to the bulk polarization},\ }\href
  {https://doi.org/10.1103/PhysRevB.103.195132} {\bibfield  {journal} {\bibinfo
   {journal} {Phys. Rev. B}\ }\textbf {\bibinfo {volume} {103}},\ \bibinfo
  {pages} {195132} (\bibinfo {year} {2021})}\BibitemShut {NoStop}%
\bibitem [{\citenamefont {Zak}(1989)}]{Zak}%
  \BibitemOpen
  \bibfield  {author} {\bibinfo {author} {\bibfnamefont {J.}~\bibnamefont
  {Zak}},\ }\bibfield  {title} {\bibinfo {title} {Berry's phase for energy
  bands in solids},\ }\href {https://doi.org/10.1103/PhysRevLett.62.2747}
  {\bibfield  {journal} {\bibinfo  {journal} {Phys. Rev. Lett.}\ }\textbf
  {\bibinfo {volume} {62}},\ \bibinfo {pages} {2747} (\bibinfo {year}
  {1989})}\BibitemShut {NoStop}%
\end{thebibliography}%

\end{document}